\documentclass[twocolumn,prd,aps,tightenlines,floats,floatfix,preprintnumbers,nofootinbib,eqsecnum]{revtex4}

\def\sp{\kern +3pt}
\def\sm{\kern -3pt}
\def\spQ{\kern +6pt}

\def\bea{\begin{eqnarray}}
\def\eea{\end{eqnarray}}

\def\sfrac#1#2{{\textstyle \frac{#1}{#2}}}

\newcommand{\ket}[1]{|#1\rangle}

\newcommand{\ms}{\scriptscriptstyle}

\def\be{\begin{equation}}
\def\ee{\end{equation}}
\def\ba{\begin{eqnarray}}
\def\ea{\end{eqnarray}}

\usepackage{graphics}
\usepackage{graphicx}
\usepackage{epsf}
\usepackage{amsmath}
\usepackage{amssymb}

\usepackage{xcolor}

\usepackage{ulem}
\begin{document}

\phantom{0}
\vspace{-0.2in}
\hspace{5.5in}

\preprint{{\bf LFTC-19-10/48}}

\vspace{-1in}

\title
{\bf Hyperon electromagnetic timelike  
elastic form factors at large $q^2$}
\author{G.~Ramalho$^1$, M.~T.~Pe\~na$^2$ and 
K.~Tsushima$^1$}
\vspace{-0.1in}

\affiliation{$^1$Laborat\'orio de 
F\'{i}sica Te\'orica e Computacional -- LFTC, \\
Universidade Cruzeiro do Sul and 
Universidade Cidade de  S\~ao Paulo, \\
01506-000,   S\~ao Paulo, SP, Brazil
\vspace{-0.15in}}
\affiliation{$^2$Centro de F\'{i}sica Te\'orica e de Part\'{i}culas (CFTP),
Instituto Superior T\'ecnico (IST),
Universidade de Lisboa,
Avenida Rovisco Pais, 1049-001 Lisboa, Portugal}

\vspace{0.2in}
\date{\today}

\phantom{0}

\begin{abstract}
We present estimates of the hyperon elastic 
form factors for the baryon octet and 
the $\Omega^-$ baryon 
for large four-momentum transfer squared, $q^2$,
in the timelike region ($q^2>0$).
Experimentally, those form factors can be extracted from the 
$e^+ e^- \to B \bar B$ and $p \bar p \to B \bar B$ processes,
where $B$ stands for a general baryon.
Our results are based on calculations 
of the elastic electromagnetic form factors 
in the spacelike region ($Q^2 = - q^2 > 0$)
within a covariant quark model. 
To connect the results in the spacelike region to those in the timelike region, 
we use asymptotic relations between 
the two regions which are constraints derived from 
analyticity and unitarity.
We calculate the effective form factors $|G(q^2)|$
and compare them with the integrated cross section data $\sigma_{\rm Born} (q^2)$ 
from BaBar, BES III, and CLEO.
The available data are at the moment restricted to $\Lambda$,
$\Sigma^0$, $\Sigma^-$, $\Xi^-$, $\Xi^0$, and $\Omega^-$
as well as to $e^+ e^- \to \Lambda \bar \Sigma^0 $ 
and $e^+ e^- \to \Sigma^0 \bar \Lambda$ reactions.
Our results provide useful reference for future experiments
and seem to indicate that the present data 
are still in the non-perturbative QCD region, 
while the onset for the asymptotic constraints from 
analyticity and unitarity happens much before  
the region of the perturbative QCD falloff of the form factors.
\end{abstract}

\vspace*{0.9in}  
\maketitle

\section{Introduction}

The understanding of internal structure 
of hadrons has been a great challenge 
after the discovery that the proton is not a pointlike particle.
In the last decades, great progress has been made 
in the study of the nucleon electromagnetic structure,
particularly through the scattering of electrons with nucleon targets
($\gamma^\ast N \to N$ transition) which probes the spacelike momentum transfer 
kinematic region 
($Q^2  \ge 0$)~\cite{NSTAR,Perdrisat07,Denig13,Pacetti15a}.
For hyperons ($B$), however, it is difficult 
to get information on the internal structure  
based on the $\gamma^\ast B \to B$ process 
due to their very short lifetimes.
The available information is restricted,
at the moment, only to the magnetic moments of a few hyperons 
(determined at $Q^2=0$).

The other possibility of disclosing the electromagnetic 
structure of baryons is $e^+ e^-$ scattering.
It enables us to access the timelike region ($q^2= -Q^2 > 0$) and
was proposed a long time ago 
by Cabibbo and Gatto~\cite{Cabibbo61a},
however, it became possible only recently.
The  $e^+ e^- \to B \bar B$ (and the inverse) reactions
open a new opportunity to study the role of valence quark effects, 
clusters of two-quark pairs (diquarks), 
and different
quark compositions~\cite{Dobbs14a,Dobbs17a,Kroll93a,Jakob93b,Jaffe03,Wilczek04,Selem06}.
The timelike region form factors appear as 
a viable tool to determine the hyperon structure, near the threshold 
as well as in the large-$q^2$ region,
where in the latter, perturbative effects are expected 
to dominate~\cite{Dobbs14a,Dobbs17a,Cabibbo61a,Aubert06,Seth13,Schonning17,Li16a,Pacetti15a}.
A significant amount of data are already
available for the proton  
($e^+ e^- \to p \bar p$)~\cite{Aubert06,Pacetti15a}.
In the present study we focus 
on the reactions involving hyperons in the final states.
Data associated with hyperon electromagnetic form factors
in the timelike region also became  available in facilities such as 
BaBar~\cite{Aubert07a}, BES-III~\cite{Ablikim18a,Ablikim13a},
and CLEO~\cite{Dobbs14a,Dobbs17a}.
The available data cover the high-$q^2$ region 
where we can expect to probe perturbative QCD (pQCD) physics.

From the theoretical side, there have been 
only a few different attempts in interpreting
the  hyperon timelike electromagnetic form factor 
data~\cite{Korner77,Haidenbauer92,Dubnickova93,Dalkarov10,Haidenbauer16,Yang18a,Cao18,Yang19}.
Although results from $e^+ e^-$ and $p \bar p$ annihilation experiments are
already available or being planned in the near future e.g. by   
the PANDA experiment at FAIR-GSI~\cite{Singh16a},
theoretical calculations of hyperon electromagnetic 
timelike form factors are scarce. 
The results presented here intend to fill that gap.

In the large-$q^2$ region one can expect 
the behavior predicted by 
pQCD~\cite{Brodsky73,Brodsky75,Lepage80,Carlson87,Belitsky03}.
However, some of the aspects from pQCD, 
including the $q^2$ dependence of the form factors,
can be seen only at very high $q^2$.
In the region covered by the present experiments, 
finite corrections for the large-$q^2$ behavior may be still relevant.

One of the goals of the present work is 
to provide calculations to
be compared with the recent experimental 
determinations of the $e^+ e^- \to B \bar B$ cross sections from 
CLEO, BaBar, and BES-III,
and to use them to guide new experiments also for larger $q^2$.
The results presented here can be used to 
study the onset of the region for the validity of 
asymptotic behavior.

Our estimates are based on results of a relativistic quark model 
for the spacelike region~\cite{Octet1,Omega}. In this work we focus on
the general properties of the integrated
cross section $\sigma_{\rm Born} (q^2)$
and the effective form factor $|G(q^2)|$ for large $q^2$.
Based on these, we test 
model-independent asymptotic relations between the 
form factors in the spacelike and the timelike regions~\cite{Pacetti15a}.
We use those relations to calculate the magnetic 
and electric form factors in the timelike region,
and give estimates 
for the  effective form factor $G(q^2)$ of the 
$\Lambda$, $\Sigma^+$, $\Sigma^0$,  $\Sigma^-$,
$\Xi^0$, $\Xi^-$ and $\Omega^-$ baryons.
An interesting aspect that emerges from our results and 
the comparison with the data is that the region of $q^2$ where 
these model-independent relations may start to hold, 
differ from the (even larger) $q^2$ region of pQCD.
This result is discussed and interpreted in terms of 
the physical scales included in our model.

In addition to the effective form factor $G(q^2)$,
we calculate also the individual form factors $|G_M|$ and $|G_E|$,
and determine their relative weights for the effective form factor.
Most existing studies  are based on the approximation $G_M \equiv G_E$, 
equivalent to $G (q^2) = G_M(q^2)$.  
However, it is important to notice that
although by definition $G_M = G_E$ 
at the threshold of the timelike region ($q^2= 4 M_B^2$,
where $M_B$ is the mass of the baryon),
there is no proof that this relation 
holds for higher values of $q^2$.
Therefore, in the present work  
we compare the result of the approximation 
$G = G_M$ with the exact result.
The difference between the two results is a measure 
of the impact of $G_E$ in the magnitude 
of the effective form factor $G$.

It is worth mentioning that, at present, calculations
of the timelike form factors based on a formulation in
Minkowski space ($q^2 =q_0^2 - {\bf q}^2$) are very important, 
since the timelike region, in practice, is  still 
out of reach of the methods as lattice QCD simulations. 
Also most of the Dyson-Schwinger-equation-based approaches,  
formulated in the Euclidean space, are still restricted 
to mass conditions compatible with singularity-free kinematic regions.
Their extension to regions where singularities 
can be crossed requires elaborate contour 
deformation techniques~\cite{Eichmann19a}.

This article is organized as follows:
In the next section we describe the general formalism associated 
with the $e^+ e^- \to B \bar B$ processes and 
their relation with the form factors $G(q^2)$.
In Sec.~\ref{secModel}, we review in detail 
the relativistic quark model used here, 
which was previously tested in calculations of 
several baryon elastic form factors 
in the spacelike region.
The model-independent relations used for the 
calculations in the large-$q^2$ region 
are discussed in Sec.~\ref{secLargeQ2}.
The numerical results for the 
timelike form factors are presented and compared  
with the experimental data in Sec.~\ref{secResults}.
The outlook and conclusions are given in Sec.~\ref{secConclusions}.
Additional details are included in the Appendixes.

\section{Formalism}
\label{secFormalism}

We start our discussion with the formalism 
associated with spin-1/2 baryons with positive parity ($1/2^+$).
In the following we represent the mass of the baryon by 
$M_B$ and use the notation  $\tau = \frac{q^2}{4 M_B^2}$.

Within the one-photon-exchange approximation
(equivalent to the impulse approximation in the spacelike region) 
one can interpret the $e^+ e^- \to B \bar B$ 
reaction as the two-step process $e^+ e^- \to \gamma^\ast \to B \bar B$,
and  the integrated cross section 
in the $e^+ e^-$ center-of-mass frame becomes~\cite{Dobbs14a,Aubert06,Haidenbauer14}
\ba
\sigma_{\rm Born} (q^2) =
\frac{4\pi \alpha^2 \beta C}{3 q^2}
\left(1  + \frac{1}{2 \tau} \right)
|G (q^2)|^2,
\label{eqSigma1}
\ea
where $G(q^2)$ is an effective form factor for the baryon
$B$ (spin 1/2 and positive parity),
$\alpha \simeq 1/137$ is the fine-structure constant,
$\beta$ is a kinematic factor defined by 
$\beta = \sqrt{1 - \frac{1}{\tau}}$,
and $C$ is a factor which depends on the charge of the baryon.
The factor $C$ is equal to 1 for neutral baryons. 
For charged baryons $C$, it takes into account the Coulomb effects 
near the threshold~\cite{Dobbs14a,Aubert06,Tzara70a,Denig13}, given by 
the Sommerfeld-Gamow factor 
$C = \frac{y}{1- \exp(-y)}$, with 
$y = \frac{\pi \alpha}{\beta} \frac{2M_B}{\sqrt{q^2}}$.
In the region of interest of the present study, 
at large $q^2$ ($\tau \gg 1$),
one has $C \simeq 1$.

The magnitude of the 
effective form factor $G$  is defined by the combination 
of the electric and magnetic form factors~\cite{Dobbs14a,Aubert06,Haidenbauer14} as
\ba
|G (q^2)|^2 &=& 
\left( 1 + \frac{1}{2 \tau} \right)^{-1}
\left[  
|G_M (q^2)|^2 + \frac{1}{2 \tau} |G_E (q^2)|^2\right], \nonumber \\
& =& 
\frac{2 \tau |G_M (q^2)|^2  + |G_E (q^2)|^2 }{2 \tau + 1}.
\label{eqGeff1}
\ea

Equations (\ref{eqSigma1}) and (\ref{eqGeff1}) 
are very useful, since they mean that one can describe 
the integrated cross section
$\sigma_{\rm Born}$ from the knowledge of 
a unique, effective function $G(q^2)$
defined by the magnetic and the electric form factors.
Note that the form factors $G_M$ and $G_E$
are complex functions of $q^2$ in the timelike region.
The relations  (\ref{eqSigma1}) and (\ref{eqGeff1}) 
are particularly practical
to calculate $\sigma_{\rm Born} (q^2)$,
because they enable us to estimate the integrated cross section
without taking into account  the relative phases 
between the form factors $G_M$ and $G_E$.

Assuming charge invariance of the electromagnetic interaction,
namely that the spacelike and timelike photon-nucleon vertices 
$\gamma pp$ and $\gamma p \bar p$ 
are the same, we can estimate the timelike form factors 
in the timelike region  
from the form factors in spacelike (SL) region 
$G_M^{\rm SL} (-q^2)$ and $G_E^{\rm SL} (-q^2)$ by applying 
the large-$|q^2|$, model-independent relations~\cite{Pacetti15a},
\ba
& &
G_M (q^2) \simeq G_M^{\rm SL} (-q^2), 
\label{eqGM2} \\
& &
G_E (q^2) \simeq G_E^{\rm SL} (-q^2),
\label{eqGE2}
\ea
and therefore restricting our results to 
the very large-$q^2$ region, where the form factors are 
real functions to fulfill the Schwarz reflection principle.
These asymptotic relations are a consequence 
of general physical and mathematical principles: unitarity as well 
as the Phragm\'en-Lindel\"{o}f theorem, which is valid 
for analytic functions (proved in Ref.~\cite{Pacetti15a}). 
They are exact in the mathematical $q^2 \rightarrow \infty$ limit,  
and they imply that 
the imaginary part of the form factors in the timelike region
goes to zero in that limit.

In the present work we use a quark model developed in the 
spacelike region~\cite{Octet1,Omega} 
to estimate the magnetic and electric form factors 
in the timelike region based on Eqs.~(\ref{eqGM2})--(\ref{eqGE2}).
The discussion on how these relations can be corrected 
for finite $q^2$ is made in Sec.~\ref{secLargeQ2}.
Deviations from those estimates may indicate that
the imaginary parts of the form factors in the considered 
timelike region cannot be neglected.

We will investigate, by comparing with the data, 
the degree of validity of those relations for finite $q^2$. 
Increasing the value of  $q^2$,
we can tentatively look for the  onset of the region 
where they may start to be a fairly good approximation.
It turns out that this happens much below  
the region where the pQCD falloff 
of the form factors starts to emerge, as our results will show.
We also provide  estimates for $q^2 > 20$ GeV$^2$ 
for comparison with future experiments.

The formalism used in the discussion of $1/2^+$ baryons
can also be extended to $3/2^+$ baryons based on 
the effective form factor~(\ref{eqGeff1}),
re-interpreting $G_M$ as a combination 
of the magnetic dipole and magnetic octupole form factors, 
and  $G_E$ as a combination of  the electric charge 
and electric quadrupole form factors~\cite{Korner77}.
The expressions associated with $G_M$ and $G_E$
for $3/2^+$ baryons are presented in Appendix~\ref{appSpin32}.
Using those expressions we calculate 
our results for the $\Omega^-$ baryon.

Before presenting the results of the extension 
of our model to the timelike region,  
we present a review of the 
covariant spectator quark model in the spacelike regime 
that sustains the application here.

\section{Covariant spectator quark model}
\label{secModel}

We restrict our study to
baryons with one or more strange quarks (hyperons).
In our estimates we use the covariant spectator quark model.
The covariant spectator quark model 
has been applied to the studies of 
the electromagnetic structure of 
several baryons, 
including nucleon, octet baryons,
and decuplet baryons (including $\Omega^-$) in the 
spacelike region~\cite{Spectator-Review,Nucleon,Octet1,Octet2,Octet3,LambdaSigma0,Omega,Omega2,NDeltaD,LatticeD,Nucleon2,Lattice,NDelta,LambdaStar,DeltaFF,Octet4,NDeltaTL}.

The model is based on three basic ingredients: 
\begin{enumerate}
\item
The baryon wave function $\Psi_B$, 
rearranged as an active quark and a spectator quark pair,
is represented in terms of the spin-flavor structure of 
the individual quarks with 
$SU_S(2) \times SU_F(3)$ symmetry~\cite{Nucleon,Omega}.
\item
By applying the impulse approximation, after integrating over the 
quark pair degrees of freedom
the three-quark system transition matrix element can be reduced 
to that of a quark-diquark system, 
parametrized by a radial wave function $\psi_B$~\cite{Nucleon,Nucleon2,Omega}. 
\item
The electromagnetic structure of the quark is 
parametrized by the quark form factors, 
$j_1$ (Dirac) and $j_2$ (Pauli) according to the flavor content,
 which encode the substructure associated with the gluons 
and quark-antiquark effects, and are parametrized using 
the vector meson dominance (VMD) mechanism~\cite{Omega,LatticeD,Lattice}.
\end{enumerate}

Concerning the first two points above,
the literature emphasizes the role of diquarks 
in the baryons~\cite{Kroll93a,Jakob93b,Jaffe03,Wilczek04,Selem06,Dobbs17a}. 
Our model, although based on a 
quark-diquark configuration, cannot be interpreted 
as a quark-diquark model in the usual sense, 
i.e.~a diquark as a pole of the 
quark-quark amplitude~\cite{Omega,Nucleon,Nucleon2}.
In our model, the  internal quark-quark motion is integrated out,
at the level of impulse approximation, 
but its spin structure signature survives~\cite{Nucleon}.
Therefore, the electromagnetic matrix element 
involves an effective quark-diquark vertex 
where the diquark is not pointlike~\cite{Nucleon2}.

Another difference between our model and the usual quark-diquark 
models is that we explicitly symmetrize in all quark pairs
applying the $SU(3)$ flavor symmetry~\cite{Octet1,Omega}.
Since it is well known that the exact $SU(3)$ 
flavor symmetry models are expected 
to fail due to the mass difference between 
the light quarks ($u$ and $d$) and the strange quarks,  
we break $SU(3)$ flavor symmetry in two levels.
We break the symmetry at the level of 
the radial wave functions by using 
different forms for those functions for systems 
with a different number of strange quarks 
($N_s=0,1,2$ for the baryon octet and $N_s=0,1,2,3$
for the baryon decuplet)~\cite{Octet1,Omega}.
We break the $SU(3)$ flavor symmetry also at the level 
of the quark current by 
considering different 
$Q^2$ dependence for the different quark sectors 
(isoscalar, isovector and strange quark components).

\subsection{Octet baryon wave functions}

The octet baryon wave functions associated 
with a quark-diquark system in the $S$-wave configuration 
can be expressed in the form~\cite{Octet1,Octet4}
\ba
\Psi_B(P,k) =
\frac{1}{\sqrt{2}}\left[
\phi_S^0 \left| M_A\right> + \phi_S^1 \left| M_S\right> 
\right] \psi_B(P,k),
\label{eqPsiB}
\ea
where $P$ ($k$) are the baryon (diquark) momentum,
$\phi_S^{0,1}$ are the spin wave functions 
associated with the components $S=0$ (scalar) 
and $S=1$ (vector)  of the diquark states, 
and $ \left| M_A\right>$, and $ \left| M_S\right> $ 
are the mixed antisymmetric and mixed symmetric flavor states
of the octet.
The explicit expressions for 
$ \left| M_A\right>$ and $ \left| M_S\right> $ and for $\phi_S^{0,1}$ 
are included in Appendix~\ref{appCSQM}.
For more details, see Refs.~\cite{Nucleon,Octet1,Octet4}.

Since the baryons are on-shell and the 
intermediate diquark in the covariant spectator model 
is taken also to be on-shell,
the radial wave functions $\psi_B$
can be written in a simple form 
using the dimensionless variable $\chi_B$:
\ba
\chi_B = \frac{(M_B-m_D)^2 - (P-k)^2}{M_B m_D}, 
\label{eqChiB}
\ea 
where $m_D$ is the diquark mass~\cite{Nucleon}.
One can now write the 
radial wave functions in the  Hulthen form,
according to~\cite{Nucleon,Octet1}: 
\ba
\psi_B (P,k) =
\frac{N_B}{m_D (\beta_1 + \chi_B)(\beta_i + \chi_B)},
\label{eqPsiBradial}
\ea
where $N_B$ is a normalization constant and
$\beta_i$ ($i=1,2,3,4$) are 
momentum-range parameters (in units $M_B m_D$).
The form of our baryon wave functions (\ref{eqPsiBradial}) 
was judiciously chosen to produce
at large-$Q^2$ the same behavior of the form factors as pQCD ~\cite{Nucleon,NDelta},
as discussed in Sec.~\ref{secOctetFF}.

In Eq.~(\ref{eqPsiBradial}) $\beta_1$ is 
the parameter which establishes the long-range scale 
($\beta_1 < \beta_2, \beta_3, \beta_4$),
common to all the octet baryons
and $\beta_i$ ($i=2,3,4$) are
parameters associated with the short-range scale,
varying with the different quark flavor contents.
The short-range scale is determined by $\beta_2$ ($N$),
$\beta_3$ ($\Lambda$ and $\Sigma$) and $\beta_4$ ($\Xi$).

The magnitudes of $\beta_i$ establish 
the shape of the radial wave function and determine
the falloff of the baryon form factors.
Heavier baryons have slower falloffs~\cite{Octet1}.
According to the uncertainty principle,
the values of the parameters $\beta_i$ ($i=2,3,4$) 
can also be interpreted in terms of the compactification in space  
of the baryons.
The relative ordering $\beta_2 > \beta_3 > \beta_4$ specifies
that $\Lambda$ and $\Sigma^{0,\pm}$ are more compact 
than the nucleon, and that $\Xi^{0,-}$ are more compact 
than $\Lambda$ and $\Sigma^{0,\pm}$.

\subsection{Electromagnetic current}
\label{secCurrent}

The contribution of the valence quarks for the transition current
in relativistic impulse approximation is expressed 
in terms of the quark-diquark wave functions $\Psi_B$ 
by~\cite{Nucleon,Octet1}
\ba
J_{B}^\mu = 3 \sum_{\Gamma} \int_k 
\overline{\Psi}_B(P_+,k) j_q^\mu \Psi_B (P_-,k),
\label{eqJimpulse}
\ea
where $j_q^\mu$ is the quark current operator, $P_+$, $P_-$ and $k$
are the final, initial and diquark momenta, respectively, 
and $\Gamma$ labels the diquark scalar and vector diquark polarizations.
The factor 3 takes into account the contributions 
associated with the different diquark pairs, 
and the integral symbol represent the covariant 
integration in the on-shell diquark momentum.

In Eq.~(\ref{eqJimpulse}) the quark current has 
a generic structure
\ba
j_q^\mu = j_1(Q^2) \gamma^\mu + j_2(Q^2) \frac{i \sigma^{\mu \nu}}{2 M_N},
\ea
where $M_N$ is the nucleon mass and $j_i$ ($i=1,2$) are
$SU(3)$ flavor operators.

The components of the quark current $j_i$ ($i=1,2$) 
can be decomposed as the sum of operators 
acting on the third quark in the $SU(3)$ flavor space
\ba
j_i(Q^2)=
\sfrac{1}{6} f_{i+} (Q^2)\lambda_0
+  \sfrac{1}{2}f_{i-} (Q^2) \lambda_3
+ \sfrac{1}{6} f_{i0} (Q^2)\lambda_s,
\nonumber \\
\label{eqJq}
\ea
where
\ba
&\lambda_0=\left(\begin{array}{ccc} 1&0 &0\cr 0 & 1 & 0 \cr
0 & 0 & 0 \cr
\end{array}\right), \hspace{.3cm}
&\lambda_3=\left(\begin{array}{ccc} 1& 0 &0\cr 0 & -1 & 0 \cr
0 & 0 & 0 \cr
\end{array}\right),
\nonumber \\
&\lambda_s \equiv \left(\begin{array}{ccc} 0&0 &0\cr 0 & 0 & 0 \cr
0 & 0 & -2 \cr
\end{array}
\right),
\label{eqL1L3}
\ea
are the flavor operators.
These operators act on the quark wave function in flavor space,
$q=  (\begin{array}{c c c} \! u \, d \, s \!\cr
\end{array} )^T$. 
The functions  $f_{i+}$, $f_{i-}$ ($i=1,2$) 
represent the quark isoscalar and isovector 
form factors, respectively, based on 
the combinations of the quarks $u$ and $d$.
The functions $f_{i0}$ ($i=1,2$) represent 
the structure associated with the strange quark.  
The explicit form for the quark form factors is 
included in Appendix~\ref{appCSQM}.

For the discussion of the results of this paper it is relevant
that the parametrization of these form factors is based on the VMD picture. 
The dressed photon-quark coupling 
is tied to the vector meson spectra.
Therefore, the isoscalar and the isovector form factors 
include contributions from the 
$\rho$ and $\omega$ mass poles in the light quark sector.
As for the strange quark form factors 
we include a dependence on the $\phi$ mass pole.
In both cases, we include also a contribution 
of an effective heavy meson with mass $2 M_N$ 
in order to take into account shorter-range effects 
in the quark current.
The parametrization of the current for the three quark sectors 
includes five parameters (coefficients of the vector meson terms)
in addition to the three quark anomalous magnetic moments.

\subsection{Model for the nucleon and decuplet baryons}

The model was first applied to the 
study of the electromagnetic structure of the nucleon.
The free parameters of the model 
(in the quark current and in the radial wave functions)
were calibrated by the electromagnetic form factor data 
for the proton and the neutron~\cite{Nucleon}.
The nucleon data are well described without an 
explicit inclusion of pion cloud contributions.

Taking advantage of the VMD form of the quark current 
and of the covariant form of the radial wave function,  
the model was extended to the 
lattice QCD regime~\cite{Lattice,LatticeD,Omega}.
This extension was performed by replacing 
the vector meson and nucleon masses in the VMD parametrization of the current
and in the baryon wave functions
by the nucleon and vector meson masses from the lattice.
This extension is valid for the region of 
the large pion masses, 
where there is a suppression of the meson cloud effects.

The extension has proved to be very successful 
in the description of the lattice QCD data for 
the nucleon and $\gamma^\ast N \to \Delta(1232)$ 
transition for pion masses above 
400 MeV~\cite{Lattice,LatticeD}. 
In the case of the $\gamma^\ast N \to \Delta(1232)$ transition 
the lattice data enabled us to fix the valence quark contribution
and after the extrapolation to the  physical pion mass limit, 
indirectly infer from the physical data 
the meson cloud effects~\cite{LatticeD,Siegert}.
The meson cloud effects was then seen to be significant 
in the case of the $\Delta(1232)$ due to the vicinity of its mass 
to the pion-nucleon threshold~\cite{NDeltaD,LatticeD,NDeltaTL}.

The formalism was later applied to all baryons of 
the decuplet using an $SU_F(3)$ extension 
of the model for the $\Delta(1232)$~\cite{NDeltaD,LatticeD,DeltaFF}, 
constrained by the scarce available lattice data for 
the decuplet baryon electromagnetic form factors
and the experimental magnetic moment of the $\Omega^-$~\cite{Omega}.
The strange quark component of the current 
and the decuplet radial wave functions were then determined 
by the fit to the data 
(lattice QCD and experimental $\Omega^-$ magnetic moment).
No meson cloud contributions were considered in this description 
of the baryon decuplet since those effects are suppressed in lattice calculations. 
Also, the only physical information on the meson cloud comes from the $\Delta(1232)$  
calibrated in the previous works~\cite{LatticeD}, and for the $\Omega^-$.
In this last case, meson cloud effects are expected to be very small, 
since  pion excitations are suppressed 
due to the content of the valence quark core
(only strange quarks) implying reduced kaon excitations 
given the large mass of the kaon~\cite{Omega,Omega2}.

The model for the $\Omega^-$ was later re-calibrated  
with the first lattice QCD calculation
of the $\Omega^-$ form factors at the physical mass point 
which we used to determine the 
electric quadrupole and magnetic octupole moments~\cite{Omega2}.

\subsection{Model for the octet baryons}
\label{secOctetFF}

Using the $SU(3)$ quark current determined 
in the studies of the nucleon and 
decuplet baryon systems,
the covariant spectator quark model 
was also extended to the octet baryon systems.
However, different from the decuplet case,
where a fair description of the data can be obtained 
based exclusively on the valence quark degrees of freedom, 
in the case of the octet, there is evidence 
that the pion cloud effects are significant~\cite{Octet4}.
Therefore, in the model for the baryon octet,
in addition to the valence quarks we consider also 
explicit pion cloud contributions based on 
the $SU(3)$ pion-baryon interaction~\cite{Octet1,Octet4}.

The valence quark contributions, regulated 
by the radial wave functions (\ref{eqPsiBradial}), 
were fixed by lattice QCD data.
The pion cloud contributions were calibrated by the physical data
(nucleon electromagnetic form factors and octet magnetic moments).
Compared to the previous studies of 
the nucleon~\cite{Nucleon}, we readjusted the values
of the momentum-range parameters $\beta_1$ and $\beta_2$
of the radial wave functions (\ref{eqPsiBradial}), 
and the quark anomalous moments $\kappa_u$ and $\kappa_d$  
in order to take into account the effects of the pion cloud.
More details can be found in Appendixes~\ref{appOctet} and \ref{appPionCloud}.

We discuss now the contributions from the valence quarks to the form factors.
From the structure for the 
quark current and radial wave functions, 
we obtain the following expressions
for the valence quark contributions to the 
octet baryon form factors:
\ba
F_{1B}(Q^2)&=&
B(Q^2) \times \nonumber \\  
& & \left[\frac{3}{2} j_1^A +
\frac{1}{2}
\frac{3 -\tau}{1+ \tau}
j_1^S - 2 \frac{\tau}{1+\tau} \frac{M_B}{M_N}
j_2^S \right],  \label{eqF1} \\
F_{2B}(Q^2) &=&
B(Q^2) \times  \nonumber \\
& &
\left[
\left(
\frac{3}{2}
j_2^A
-\frac{1}{2} \frac{1-3\tau}{1+\tau} j_2^S \right) \frac{M_B}{M_N}
-2 \frac{1}{1+\tau} j_1^S
\right] , \nonumber \\
& &
\label{eqF2}
\ea
with $\tau=\sfrac{Q^2}{4M_B^2}$, and
\be
B(Q^2)= \int_k \psi_B(P_+,k) \psi_B(P_-,k),
\label{eqB}
\ee
the overlap integral between
the initial and final scalar wave functions.
The function $B(Q^2)$ is independent 
of the diquark mass~\cite{Nucleon}.

The coefficients $j_i^{A,S}$ ($i=1,2$) are 
combinations of the quark form factors 
dependent on the baryon quark content.
The explicit expressions are included 
in Appendix~\ref{appCSQM}.
One concludes that the results 
are an interplay of both the structure of
the quark form factors and of the radial wave functions.

The results for the electric and magnetic 
form factors are then determined by
\ba
G_{EB} = F_{1B} - \tau F_{2B}, 
\hspace{.5cm}
G_{MB} = F_{1B} + F_{2B}.
\ea

The asymptotic behavior of the form factors $G_E$ and $G_M$ 
is determined by the asymptotic results for 
$F_{1B}$ and $F_{2B}$ from Eqs.~(\ref{eqF1})--(\ref{eqF2}).
The terms between brackets depend only 
on the quark form factors and for large $Q^2$,
and contribute to $F_{1B}$ and $Q^2 F_{2B}$ with constants.
As a consequence, the results for 
$G_E$ and $G_M$ are determined at very large $Q^2$ 
by the function $B(Q^2)$,
which in turn exclusively depends on the radial wave functions and their overlap. 
In Ref.~\cite{NDelta} it was shown that, 
if we use the radial wave functions (\ref{eqPsiBradial}),
one has $B \propto 1/Q^4$ plus logarithmic corrections.
We can conclude then that 
the combination of the quark current with the 
radial structure induces falloffs for  
the form factors consistent
with the power law of pQCD result: 
$G_E \propto 1/Q^4$ and $G_M \propto 1/Q^4$, 
in addition to logarithmic 
corrections~\cite{Brodsky75,Lepage80,Belitsky03,NDelta}.

The deviations of our results  from
the simple power law $1/Q^4$ are originated 
by contaminations from  logarithmic corrections,
or from the difference of the  quark form factors 
from their asymptotic result
($j_1^{A,S}$, $Q^2 j_2^{A,S} \to$ constant). 
The latter is regulated by large momentum scales 
associated with the VMD parametrization. 
But the different momentum falloff tails of the baryon form factors 
also play a role and relate to the difference 
in the flavor content of the constituent valence quarks 
described by the wave functions, 
as well as the VMD structure of the quark form factors.

As mentioned above, an accurate description 
of the electromagnetic structure of the octet baryons 
is achieved when we include an explicit parametrization of 
the pion cloud contributions~\cite{Octet1,LambdaSigma0}. 
The consequence of the introduction of the pion cloud effects 
is that the transition form factors 
(\ref{eqF1})--(\ref{eqF2}) have additional contributions,
which can be significant below $Q^2 <$ 2 GeV$^2$, 
and that the two contributions have to be normalized 
by a global factor $Z_B < 1$ ($\sqrt{Z_B}$ is the factor associated 
with each wave function).

In the large-$Q^2$ region, the pion cloud contributions
are suppressed and the form factors are then reduced to 
\ba
G_{EB} \to Z_B G_{EB}, \hspace{.5cm}
G_{MB} \to Z_B G_{MB},
\label{eqZGEGM}
\ea
where $G_{EB}$ and $G_{MB}$ on the r.h~s.~represent 
the valence quark estimate.
From Eq.~(\ref{eqZGEGM}), we conclude that the  
pion cloud dressing affects only the 
normalization of the form factors at large $Q^2$.
The normalization factor $Z_B$ depends only on a parameter associated 
with the pion cloud parametrization:
the parameter which determines the pion cloud 
contribution to the proton charge ($Z_N$).
In Appendix~\ref{appCSQM}, we show that 
all normalization factors can be determined by 
the normalization of the nucleon wave functions $Z_N$. 
The values of $Z_B$ (between 0.9 and 1)  
are also presented in Appendix~\ref{appCSQM}.

Our calculations for the baryon octet presented in the next sections
for the $Q^2 > 5$ GeV$^2$ region depend essentially on seven parameters: 
four momentum-range parameters ($\beta_i$),
two anomalous magnetic moments ($\kappa_u$ and $\kappa_d$) 
and one pion cloud parameter associated with the normalization 
of the octet baryon wave functions.
The parametrization for the quark current was 
determined previously in the studies 
of the nucleon and baryon decuplet systems.

\section{Model-independent relations in the large-$q^2$ regime}
\label{secLargeQ2}

In the present work we test the results of extrapolating the parametrizations 
in the spacelike region ($q^2 = -Q^2 < 0$) 
to the timelike region ($q^2 > 0$).
The calculation in the timelike region is based also 
on the model-independent relations~(\ref{eqGM2})--(\ref{eqGE2}) 
results for large $Q^2$.


Concerning the relations~(\ref{eqGM2})--(\ref{eqGE2}),
they map the region of $q^2$: $]-\infty,0]$
into the region  $[0, + \infty[$.
Note, however, that $q^2=0$ is not the center point 
of the reflection symmetry  that relates timelike 
and spacelike regions, because of the unphysical gap region 
$]0,4 M_B^2[$ between them.
The reflection symmetry center point between the two regions lies, in fact,
inside this interval and can be tentatively taken 
as $q^2=2M_B^2$ instead of $q^2=0$.
This consideration leads us to correct the relations 
(\ref{eqGM2})--(\ref{eqGE2}) by introducing finite corrections to $q^2$,
\ba
& &
G_M(q^2)  \simeq G_M^{\rm SL} (2M_B^2 -q^2),
\label{eqGMnew} \\
& &
G_E(q^2)  \simeq G_E^{\rm SL} (2M_B^2 -q^2). 
\label{eqGEnew} 
\ea
While the difference between using  (\ref{eqGM2})--(\ref{eqGE2})
and (\ref{eqGMnew})--(\ref{eqGEnew})
is naturally negligible for very large $q^2$, and is
immaterial in the mathematical $q^2\rightarrow \infty$ limit, it can be 
non-negligible otherwise.
In the next section we check that this is indeed the case when
one gets to values in the range $q^2=10$--20 GeV$^2$.

In the calculations presented in the next section, 
Eqs.~(\ref{eqGMnew})--(\ref{eqGEnew}) provide a central value 
for our results of the form factors,
Eqs.~(\ref{eqGM2})--(\ref{eqGE2}), a lower limit,
while the estimate where we replace 
$G_l^{\rm SL} (-q^2)$ by $G_l^{\rm SL} (4M_B^2 -q^2)$ ($l=M,E$) 
gives the upper limit.

An important point that is addressed in the next section 
is to know how far 
the region of the asymptotic relations (\ref{eqGMnew})--(\ref{eqGEnew})
is from the pQCD 
region characterized by the relations 
$G_M \propto 1/q^4$ and $G_E \propto 1/q^4$~\cite{Brodsky75,Lepage80,Belitsky03}.

\section{Results}
\label{secResults}

In this section we present the results in the timelike region for 
the $\Lambda$,  $\Sigma^-$, 
$\Sigma^0$, $\Xi^-$ and $\Xi^0$ of the  baryon octet  and also for the 
$\Omega^-$ (baryon decuplet).
The results for the baryon octet are based on the model 
from Ref.~\cite{Octet1}.
The results for the $\Omega^-$  
are based on the model from Ref.~\cite{Omega2}.

\subsection{Octet baryons}

\begin{figure}[t]
\centerline{\vspace{0.5cm}  }
\centerline{
\mbox{
\includegraphics[width=3.1in]{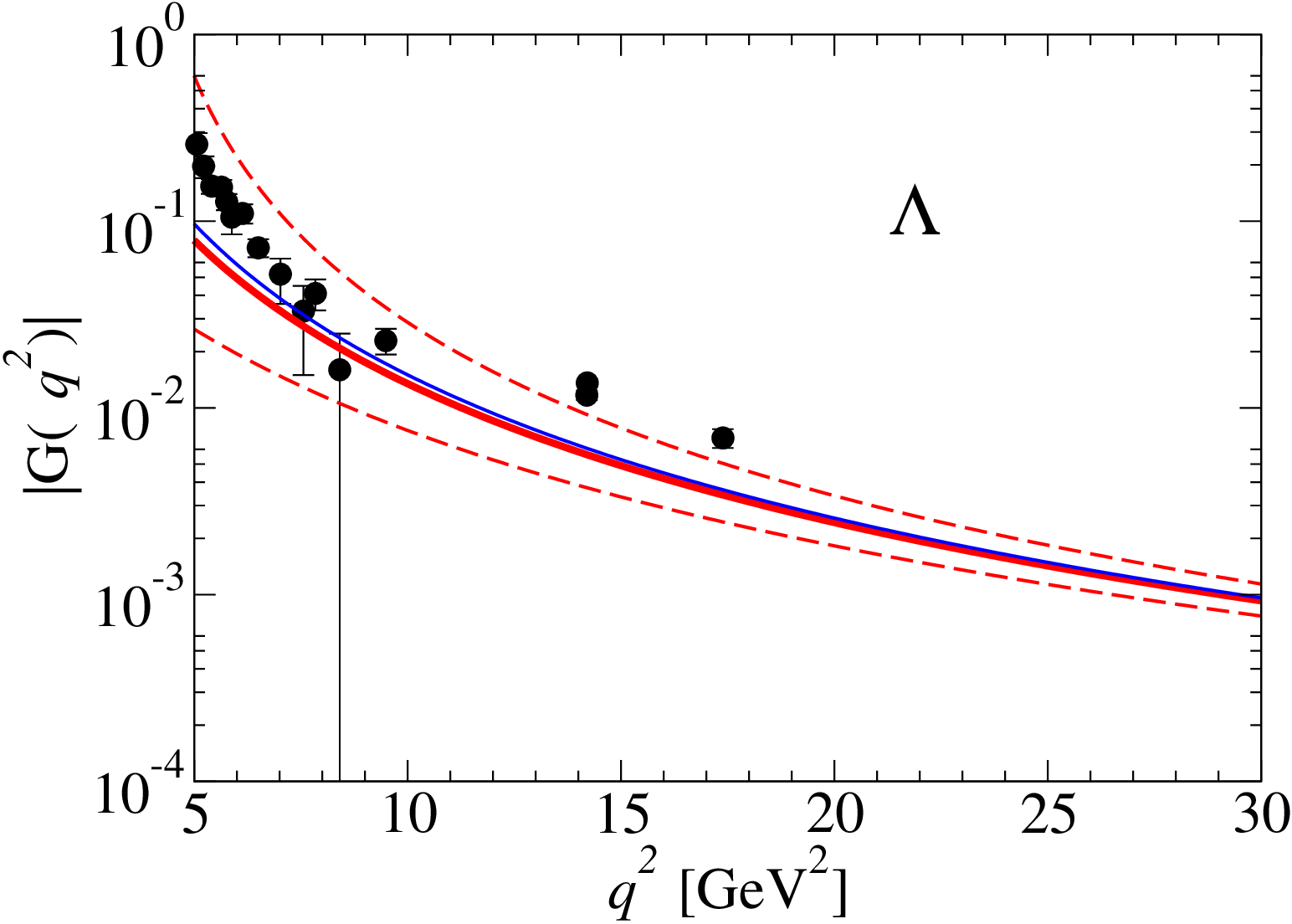} 
}}
\caption{\footnotesize{
Timelike form factor $G$ for the $\Lambda$.
Data are from Refs.~\cite{Aubert07a,Dobbs14a,Dobbs17a,Ablikim18a}. 
The thick solid line is based on Eqs.~(\ref{eqGMnew})--(\ref{eqGEnew}).
The dashed lines represent the upper limit 
$G_l (q^2) = G_l^{\rm SL} (4 M_B^2 -q^2)$ 
and the lower limit $G_l (q^2) = G_l^{\rm SL} (-q^2)$ ($l=M,E$).
The thin solid line is obtained with the approximation $G_E=G_M$.}}
\label{figLambda}
\end{figure}
\begin{figure*}[t]
\centerline{\vspace{0.5cm}  }
\centerline{
\mbox{
\includegraphics[width=3.1in]{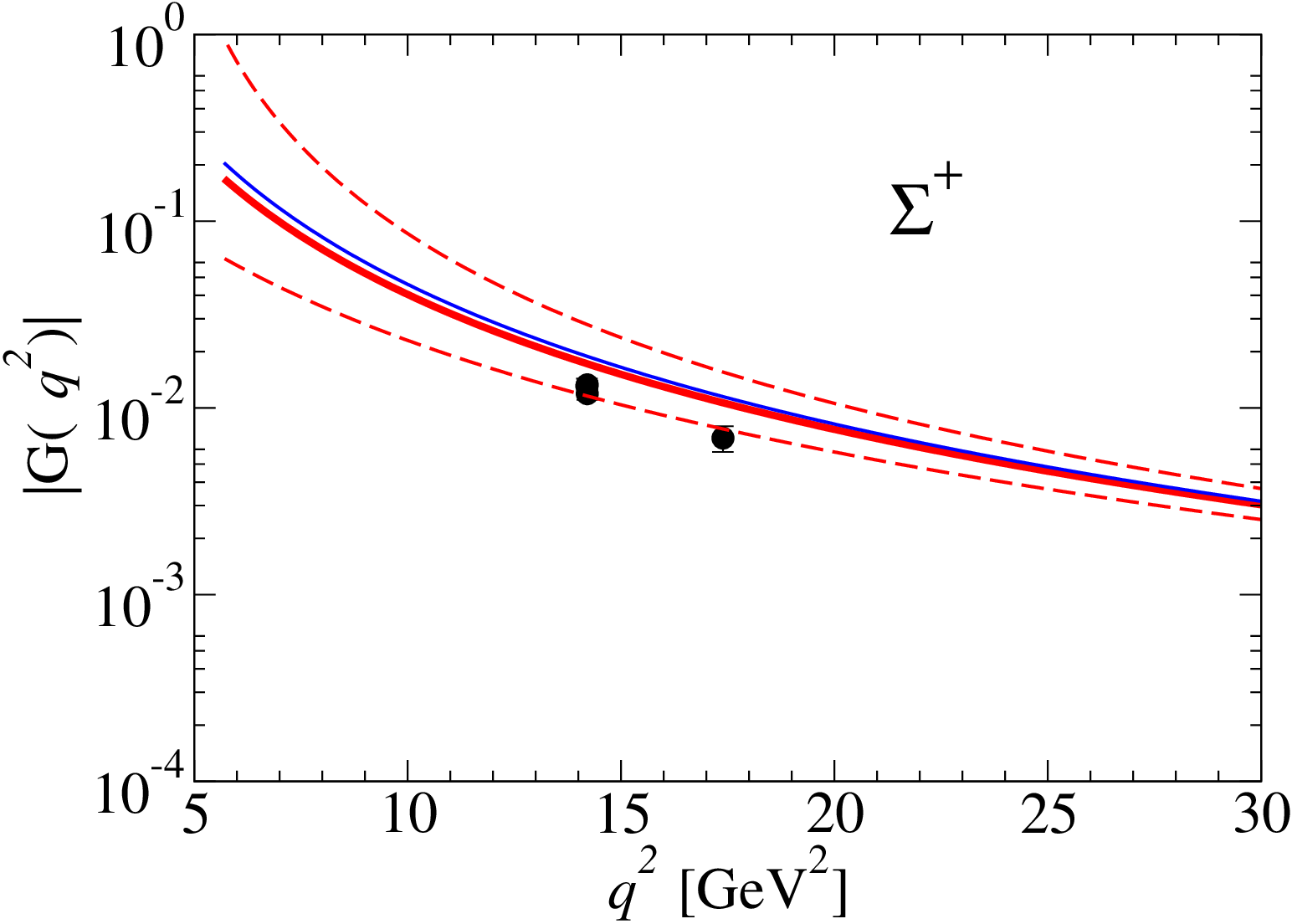} \hspace{.6cm}
\includegraphics[width=3.1in]{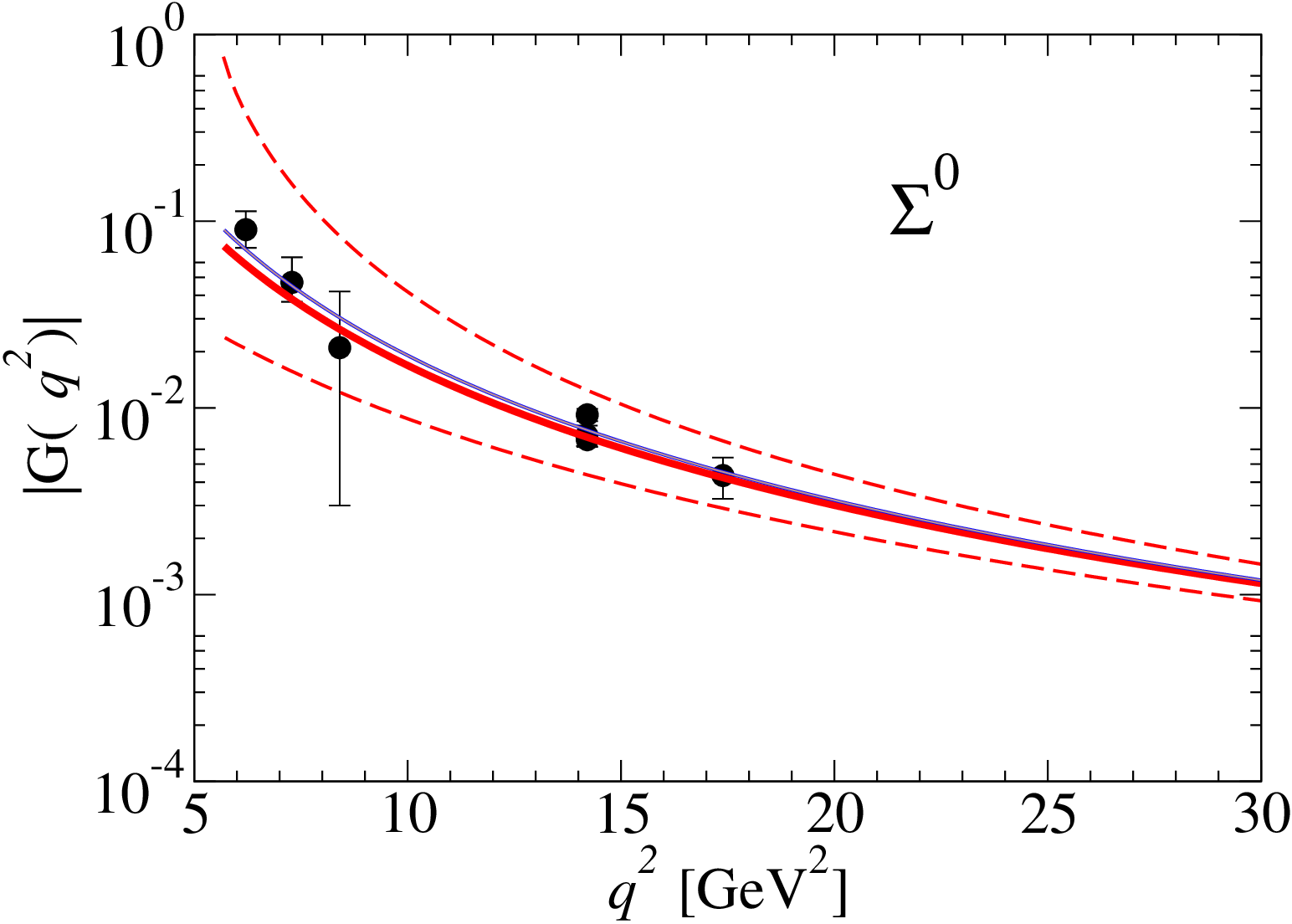} }}
\caption{\footnotesize{
Timelike form factor $G$ for the $\Sigma^+$ 
(left) and $\Sigma^0$ (right).
Data are from Refs.~\cite{Aubert07a,Ablikim13a,Dobbs14a,Dobbs17a}. 
See also caption of Fig.~\ref{figLambda}. 
}}
\label{figSigma}
\end{figure*}
\begin{figure*}[t]
\centerline{\vspace{0.5cm}  }
\centerline{
\mbox{
\includegraphics[width=3.1in]{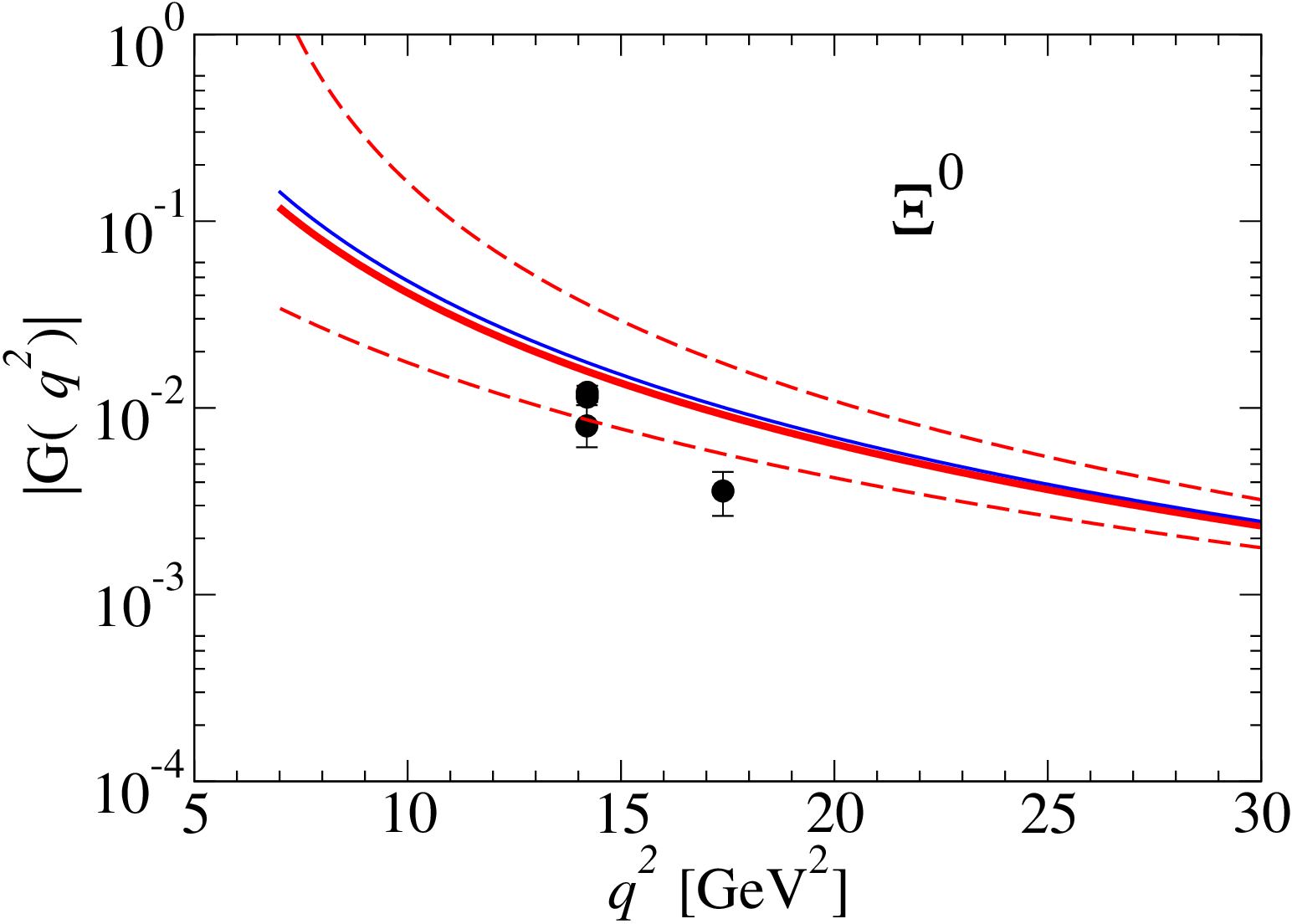} \hspace{.6cm}
\includegraphics[width=3.1in]{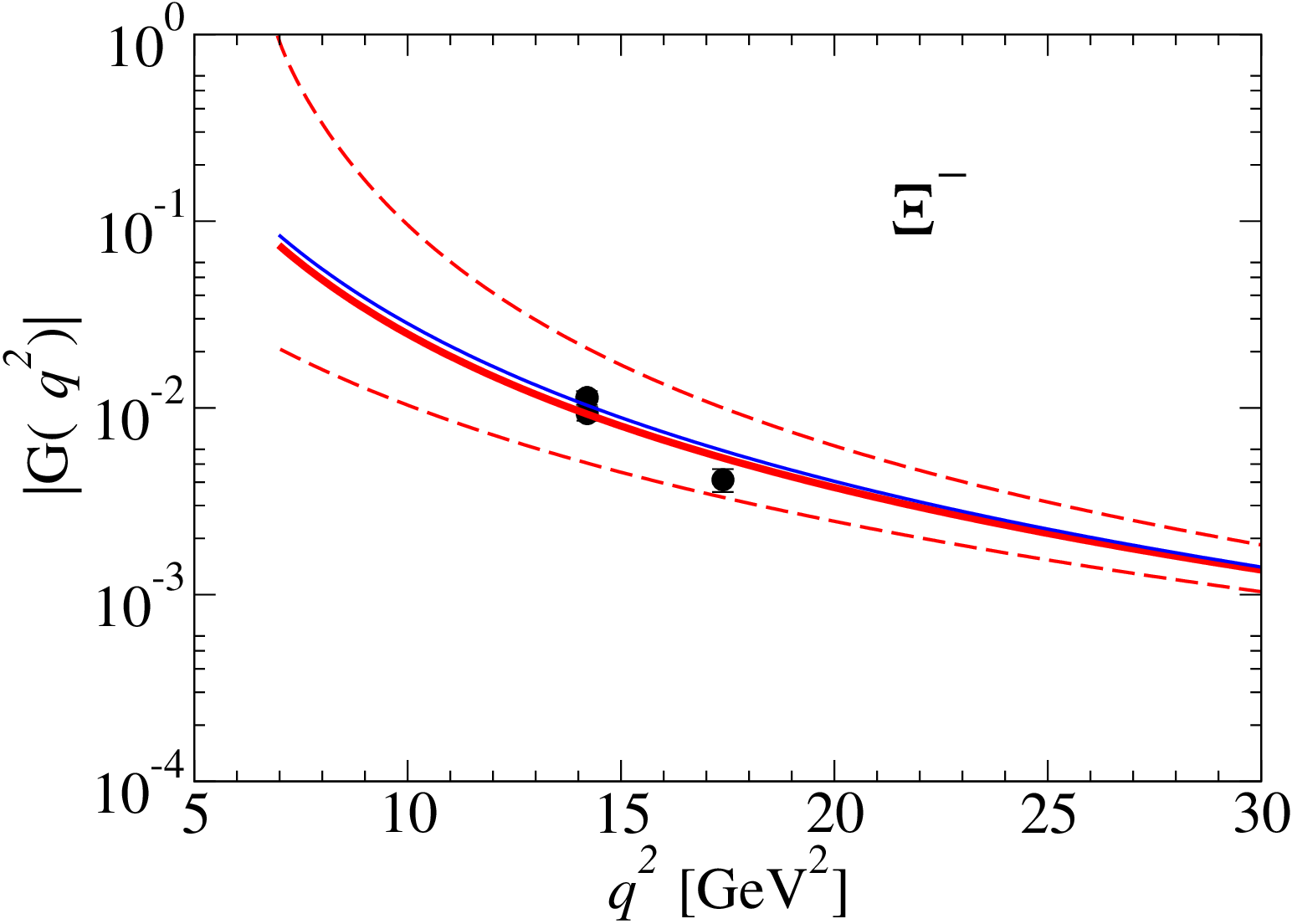} }}
\caption{\footnotesize{
Timelike form factor $G$ for the $\Xi^0$ 
(left) and $\Xi^-$ (right).
Data are from Refs.~\cite{Ablikim13a,Dobbs14a,Dobbs17a}.
See also caption of Fig.~\ref{figLambda}. 
}}
\label{figXi}
\end{figure*}

The results of our model in the timelike region
are presented in Figs.~\ref{figLambda},  \ref{figSigma} 
and \ref{figXi} for the cases of $\Lambda$, $\Sigma^-$, 
$\Sigma^0$, $\Xi^-$ and $\Xi^0$. 
The thick solid lines represent our best estimate 
based on Eqs.~(\ref{eqGMnew})--(\ref{eqGEnew}).
The dashed lines represent the upper limit 
$G_l (q^2) = G_l^{\rm SL} (4 M_B^2 -q^2)$,  
and the lower limit $G_l (q^2) = G_l^{\rm SL} (-q^2)$ ($l=M,E$).
The thin solid line results are those obtained with the approximation 
$G_E=G_M$, and will be discussed later.
Naturally, all curves get closer together as $q^2$ increases.
In all cases, we use the experimental masses or 
the averages (respectively for $\Sigma$ and $\Xi$). 
We recall that in the present model 
the  $SU(3)$ flavor symmetry is broken 
by the radial wave functions and that 
the quark electromagnetic structure is  
parametrized based on a VMD representation.

Our estimates are compared with the world data for the hyperon 
electromagnetic form factors in the timelike region.
The data  for the $\Lambda$, $\Sigma^0$, and $\Lambda \bar \Sigma^0$
(from $e^+ e^- \to \Lambda \bar \Sigma^0$
and $e^+ e^- \to \Sigma^0 \bar \Lambda$ reactions)
for values of 
$q^2$ up to 9 GeV$^2$, are from BaBar~\cite{Aubert07a}.
There are also data from BES-III
for the $\Lambda$~\cite{Ablikim18a} 
below $q^2=10$ GeV$^2$ and  
for $\Sigma^0$, $\Sigma^+$, $\Xi^-$ and $\Xi^0$
for $q^2 \simeq 14.2$ GeV$^2$ [$\psi(3770)$ decay]~\cite{Ablikim13a}.
Finally, there are data from 
CESR (CLEO-c detector)~\cite{Dobbs14a,Dobbs17a}
for the baryon octet
($\Lambda$, $\Lambda \bar \Sigma^0$, 
$\Sigma^0$, $\Sigma^+$, $\Xi^-$, and $\Xi^0$) and $\Omega^-$
for $q^2 \simeq 14.2$ and 17.4 GeV$^2$
[$\psi(3770)$ and $\psi(4170)$ decays].
In the near future, 
we expect results on the proton-antiproton scattering from PANDA
($p \bar p \to B \bar B$)~\cite{Singh16a}.

Contrary to the case of the proton form factor data
in the timelike region, which is about 2 times 
larger than those in 
the spacelike region~\cite{Kroll93a,CLOE05,Fermilab93,Fermilab99}, 
the hyperon form factors have about the same  
magnitude (central value lines in the figures)  
in both regions (spacelike and timelike).
Our results suggest that the available data 
may already be within the asymptotic region 
where Eqs.~(\ref{eqGMnew})--(\ref{eqGEnew}) are valid,
with the deviations consistent with a variation of the argument of $G$
from $q^2$ (lower limit) up to $q^2 -4M_B^2$ (upper limit),  
denoting that the reflection center point is
within the unphysical region.
In the model of Ref.~\cite{Kuraev12} this seems also to be the case.


From the figures, we can conclude that 
our estimates (central values) are close to the data 
for $q^2 > 8$ GeV$^2$ in most cases.
For the $\Lambda$ case our results underestimate the data.
For the $\Sigma^+$ and $\Xi^0$ cases our results overestimate the data.
However, in general, our results are reasonably close to the data.
To compare our results with the data for
larger values of  $q^2$ ($q^2 \simeq 14.2$ and 17.4 GeV$^2$ 
from CLEO-c~\cite{Dobbs14a,Dobbs17a}), we show in Table~\ref{tabAverage} 
the average ratios of the experimental values and our estimates.
Note that for the $\Xi^0$  we have an underestimate 
of 40\% ($\approx 0.6$) and for the $\Lambda$
an overestimate of more than 100\% ($\approx 2.2$).
This feature is similar to the proton case.
When we average in the baryon indices, however,  
we obtain a ratio of 1.12, meaning that the baryon average value 
is very close to our model estimate.

\begin{figure}[t]
\centerline{\vspace{0.5cm}  }
\centerline{
\mbox{
\includegraphics[width=3.1in]{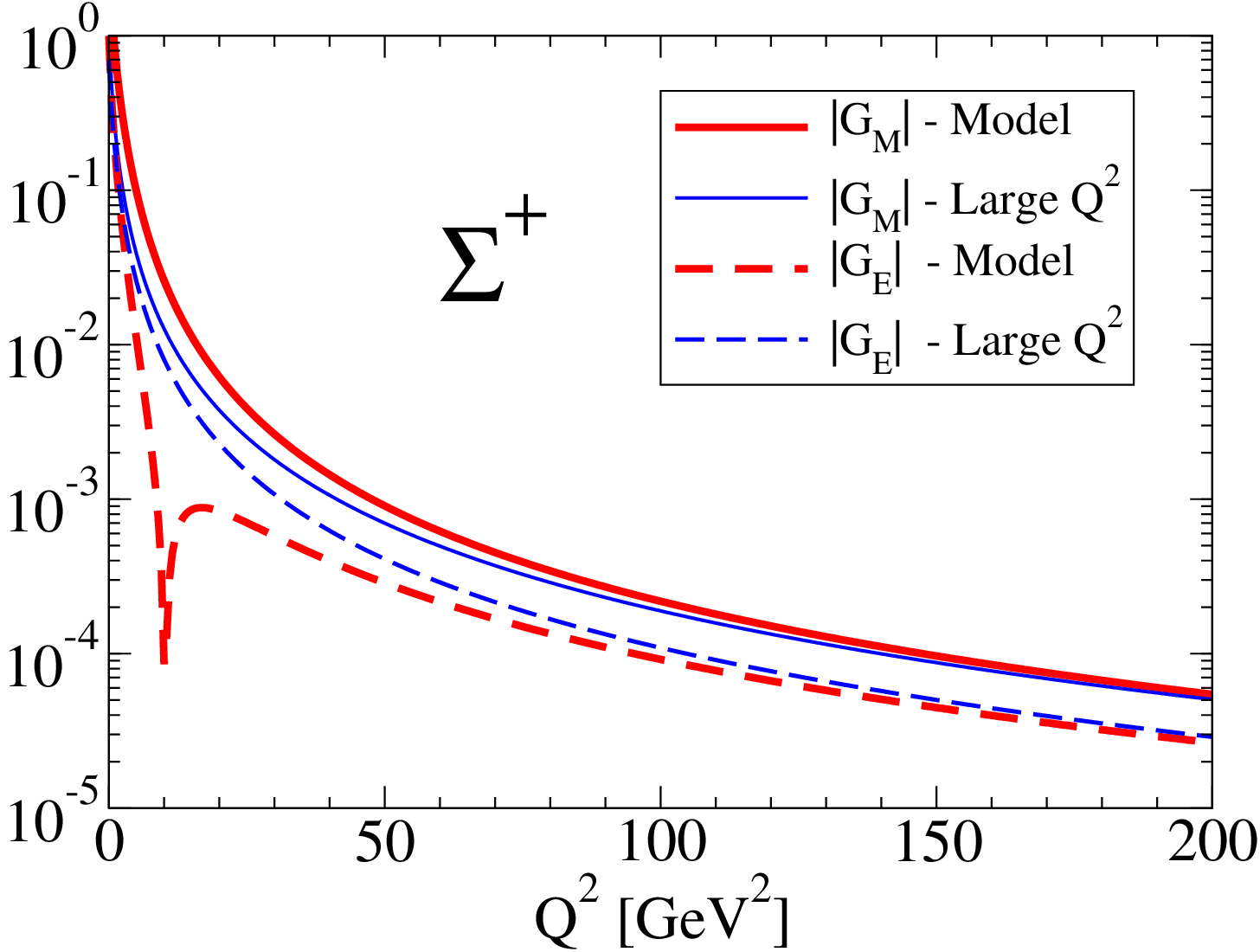} }}
\caption{\footnotesize{
$\Sigma^+$ form factors. 
Comparison between the results for 
$|G_M|$ and  $|G_E|$ for a model with 
an exact quark current (Model) 
and the results where we consider only 
the leading order term in  
$Q^2$ for the quark current (Large $Q^2$).
}}
\label{figX}
\end{figure}

Although our results corroborate 
the idea that the region shown is close to the  region
where the relations~(\ref{eqGMnew})--(\ref{eqGEnew}) hold,
that does not mean that the region is close to the pQCD region, 
where $G_E \propto 1/q^4$  and $G_M \propto 1/q^4$.
Calculations in the spacelike region
where we consider the leading order term 
of the asymptotic quark current suggest that 
the first signs of the pQCD behavior 
$G_E \propto 1/Q^4$ and $G_M \propto 1/Q^4$ (with log corrections) 
appear only for  $q^2  \approx 100$ GeV$^2$.
An example of the convergence for $|G_M|$ and  $|G_E|$ 
to the perturbative regime is presented in Fig.~\ref{figX} 
for the case of $\Sigma^+$. 
The lines with the label ``Model'' indicate the exact 
result; the lines with the label ``Large $Q^2$'' 
indicate the calculation with the asymptotic quark current. 
Similar behavior can be observed for the other hyperons.

The sharp minimum on $|G_E|$ is a consequence of
the zero crossing for $Q^2 \simeq 10$ GeV$^2$ 
($G_E$ becomes negative above that point).
This case is similar to the case of the proton,
where there is the possibility of a zero-crossing 
near  $Q^2 \simeq 9$ GeV$^2$, according to 
recent measurements at Jefferson Lab~\cite{Puckett17}.
The zero crossing is also expected for other hyperons.

That the leading order pQCD behavior of the form factors only appears for 
higher $q^2$, can be interpreted as the interplay between 
the meson masses that enter the model through the constituent 
quark electromagnetic form factors
(describing the photon-quark coupling) and the tail of  
the baryon wave functions that enter the overlap integral.
On one hand, the quark electromagnetic form factors carry information 
on the meson spectrum, being  parametrized 
using the VMD mechanism in our model.
Depending on the hyperon flavor, one has different contributions from the poles 
associated with light vector mesons (0.8--1.0 GeV)
and an effective heavy vector meson (1.9 GeV).
Those vector meson masses provide a natural scale, 
which regulates the falloff of the hyperon electromagnetic form factors.
Note that the light vector meson masses (0.8--1.0 GeV)
correspond to a large scale compared to the low-$Q^2$ scale 
of QCD ($\sim 0.3$ GeV) and of the constituent quark masses.
On the other hand, the momentum falloff tail of
the wave functions of the heavy baryons 
is associated with larger cutoffs.
The same effect is observed in lattice QCD simulations, 
where form factors associated with larger pion masses 
exhibit slower falloffs in $Q^2$ ~\cite{Alexandrou06,Lattice,LatticeD}.

For a detailed comparison with the present and future data, 
we present in Tables~\ref{tabSigma} and~\ref{tabXi}
our estimates for $G$ at larger values of $q^2$.
Note in particular that we present predictions for $\Sigma^-$,
a baryon for which there are no data at the moment.
The results in the tables can be used to 
calculate the ratios between the form factors 
associated with different baryons.

\begin{table}[t]
\begin{center}
\begin{tabular}{c  c }
\hline
\hline
   $B$  & $\left< \frac{G^{\rm exp}}{G^{\rm mod}} \right>$ \\
\hline
\hline
 $\Lambda$   & 2.19 \\
 $\Sigma^+$  &  0.65 \\
 $\Sigma^0$ & 1.08 \\
 $\Xi^-$    & 1.08 \\
 $\Xi^0$    & 0.60 \\
\hline  
Average            & 1.12   \\
\hline
\hline
\end{tabular}
\end{center}
\caption{\footnotesize
Comparison between the ratios between the experimental value ($G^{\rm exp}$) 
and the model estimate of $G$ ($G^{\rm mod}$) for the different baryons,
for $q^2 \simeq 14.2$ and 17.4 GeV$^2$~\cite{Dobbs14a,Dobbs17a}. 
The last line indicates the average of all baryons.}
\label{tabAverage}
\end{table}

From the previous analysis, we can conclude 
that the effective form factor $G$ for most of the octet baryons 
with strange quarks (hyperons) 
is well described by our approximated $SU_F(3)$ model
combined with the asymptotic relations (\ref{eqGMnew})--(\ref{eqGEnew}),
since the data lie within the upper and lower limits of 
the theoretical uncertainty.

\subsubsection*{Discussion}

In the literature, there are a few estimates 
of hyperon form factors based 
on vector meson dominance~\cite{Korner77,Dubnickova93}.
The first calculation (1977)~\cite{Korner77} was 
performed with no adjustable parameters, 
before the first measurements (Orsay 1990)~\cite{DM2}.
Those estimates differ from the recent measurements by 
an order of magnitude~\cite{Dobbs14a,Dobbs17a}.
An improved VMD estimate (1993)~\cite{Dubnickova93}
gave results closer to the $\Lambda$ data
under the condition $G_M=G_E$~\cite{Seth13,Dobbs17a}.
There are also recent estimates for the 
$\Lambda$ and $\Sigma^0$ form factors based 
on phenomenological parametrizations 
of the baryon-antibaryon interaction~\cite{Dalkarov10},
asymptotic parametrizations 
and vector meson dominance parametrizations 
of the form factors~\cite{Yang18a,Cao18,Yang19}.

In our model, the $SU_F(2)$ symmetry is broken at the quark level 
since we use different parametrizations for the isoscalar
and isovector quark form factors. The dependence on the isovector component 
is more relevant for the case of the neutron
for which there are almost 
no data available~\cite{Denig13,Achasov14,Antonelli98}, 
and for the 
$e^+ e^- \to \Lambda \bar \Sigma^0 $
and $e^+ e^- \to  \Sigma^0 \bar \Lambda$
reactions, which we discuss at the end of the present section.

We now discuss the difference in magnitude between the 
electric and magnetic form factors of the octet baryon members.
The absolute value of the magnetic form factor $|G_M|$ 
is represented in Figs.~\ref{figLambda} to 
\ref{figXi} by the thin solid line, 
which is, with no exception, 
just a bit above the central (thick solid line).
Those results mean that the magnetic form factor is larger 
than the electric form factor
($|G_E| < |G_M|$)
for $\Lambda$, $\Sigma^+$, $\Sigma^0$, $\Xi^0$ and $\Xi^-$.
This conclusion is a consequence of 
the definition of $|G(q^2)|^2$ given by Eq.~(\ref{eqGeff1}).
If we express $|G_E|$ in terms of the ratio
$\alpha_{\ms G} = \frac{|G_E|}{|G_M|}$, we obtain 
$|G|^2 = |G_M^2|\left( 1  + \frac{\alpha_{\ms G}^2 -1}{1 + 2 \tau}\right)$.
Since the thick solid line is the result for the full $|G(q^2)|$ function, 
and the thin solid line is the result from assuming $|G(q^2)|=|G_M (q^2)|$, 
we conclude that although $|G_E| < |G_M|$,
the two form factors have similar magnitudes.

Our model can also be applied for the $\Lambda \bar \Sigma^0$
and $\bar \Lambda \Sigma^0$ form factors
($e^+ e^- \to \Lambda \bar \Sigma^0$ 
and $e^+ e^- \to \bar \Lambda \Sigma^0$ reactions).
It is important to notice, however, 
that the analysis of the 
$e^+ e^- \to \Lambda \bar \Sigma^0$ 
and the $e^+ e^- \to \bar \Lambda \Sigma^0$ reactions
is a bit more intricate 
than the analysis for the $e^+ e^- \to B \bar B$ reactions associated 
with the elastic form factors.
In this case there are two possible final states
($\Lambda \bar \Sigma^0$ and $\bar \Lambda \Sigma^0$).
From the experimental point of view, this
implies that the background subtraction in the 
cross section analysis is also more complex due to the proliferation
of decay channels, including the $\Lambda$ and $\Sigma^0$ decays
and the decays of the corresponding antistates.

From the theoretical point of view the 
$\gamma^\ast \Lambda \to \Sigma^0$ transition form factors 
in the spacelike region are difficult to test due 
to the lack of experimental data:
There are no experimental constraints for the electric 
and magnetic form factors, except for the transition magnetic moment.
We do not discuss here in detail our results 
for the $\gamma^\ast \Lambda \to \Sigma^0$ 
transition form factors, due to the experimental 
ambiguities and also because the main focus of this work
is the octet baryon electromagnetic form factors.
Still, we mention that we predict the 
dominance of the meson cloud contributions 
for $G_E$ and of the valence quark contributions 
for $G_M$~\cite{LambdaSigma0}.
At large $Q^2$, the magnetic form factor dominates 
over the electric form factor.
This dominance is then {\it mirrored}  to the timelike region.
Our estimate of $G$ in the timelike region   
overestimates the data by about an order of magnitude,
suggesting that the magnetic form factor dominance 
is not so strong in the timelike region.
Another interesting theoretical aspect related 
to the  $\gamma^\ast \Lambda \to \Sigma^0$ transition 
is its isovector character.
This property can be studied in the near future 
once accurate timelike data for the neutron become available
at large $q^2$.
From the combination of proton and neutron data, 
we can determine the isovector component of the nucleon form factors.
Then those can be used to study the 
$\gamma^\ast \Lambda \to \Sigma^0$ transition form factors.

\vspace{.3cm}
\begin{table}[t]
\begin{center}
\begin{tabular}{c | c   c   c }
\hline
\hline
    $q^2$ (GeV$^2$) &\hspace{.3cm} $\Sigma^+$ \hspace{.3cm} 
& \hspace{.3cm} $\Sigma^0$ \hspace{.3cm} 
& \hspace{.3cm} $\Sigma^-$  \hspace{.3cm}\\
\hline
\hline
 10  &   40.5  &  16.8  &  10.7\\
 15  &   15.1  &  6.09  &  4.12\\
 20  &   7.68  &  3.01  &  2.19 \\
 25  &   4.58  &  1.76  &  1.36 \\
 30  &   3.03  &  1.15  &  0.923 \\
 35  &   2.14  &  0.803  &  0.667\\
 40  &   1.60  &  0.592  &  0.503\\
 45  &   1.24  &  0.453  &  0.393\\
 50  &   0.980 &  0.358  &  0.315\\
 55  &   0.799 &  0.290  &  0.260 \\
 60  &   0.663 &  0.239  &  0.216\\
\hline
\hline
\end{tabular}
\end{center}
\caption{\footnotesize
Estimates for the $\Sigma$ effective form factor $G$
in units $10^{-3}$.}
\label{tabSigma}
\end{table}
\vspace{.3cm}
\begin{table}[t]  
\begin{center}
\begin{tabular}{c | c   c   c }
\hline
\hline
    $q^2$ (GeV$^2$) & \hspace{.3cm} $\Lambda$  \hspace{.3cm} 
&  \hspace{.3cm} $\Xi^0$  \hspace{.3cm}  
&   \hspace{.3cm} $\Xi^-$  \hspace{.3cm} \\
\hline
\hline
 10  &  13.4  &  41.4  &  24.9 \\
 15  &  4.90  &  13.6  &   7.99 \\
 20  &  2.43  &  6.41  &   3.75  \\
 25  &  1.43  &  3.65  &   2.13 \\
 30  &  0.927  & 2.33  &   1.36 \\
 35  &  0.648  & 1.61  &   0.933 \\
 40  &  0.476  & 1.17  &   0.679 \\
 45  &  0.365  & 0.893 &   0.514 \\
 50  &  0.288  & 0.700 &   0.402\\
 55  &  0.233  & 0.564 &   0.323 \\
 60  &  0.192  & 0.463 &   0.264 \\
\hline
\hline
\end{tabular}
\end{center}
\caption{\footnotesize
Estimates for the $\Lambda$  and $\Xi$ effective form factor $G$
in units $10^{-3}$.}
\label{tabXi}
\end{table}

\subsection{$\Omega^-$ form factors}

\begin{figure*}[t]
\centerline{\vspace{0.5cm}  }
\centerline{
\mbox{
\includegraphics[width=3.1in]{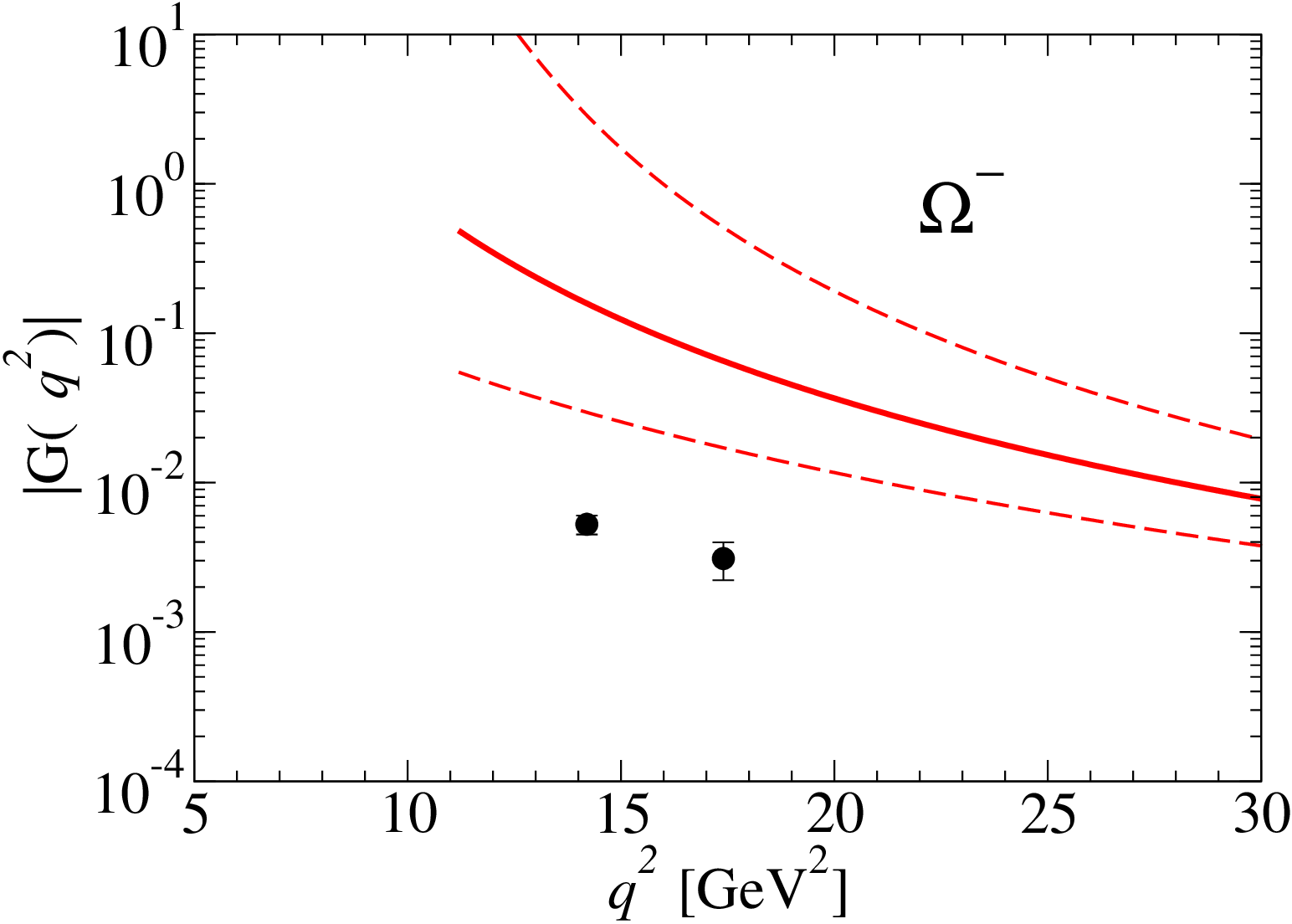} \hspace{.6cm}
\includegraphics[width=3.1in]{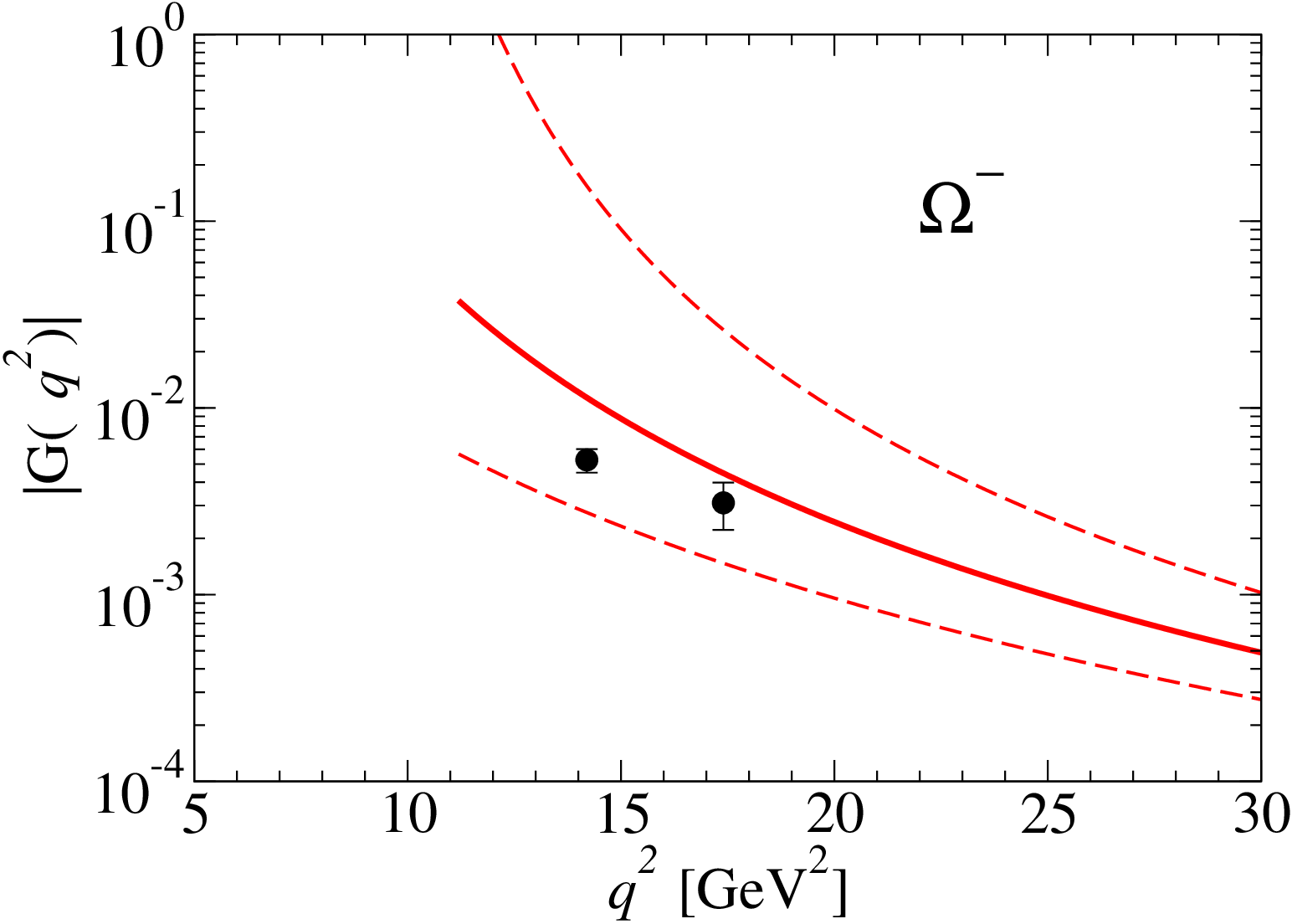} }}
\caption{\footnotesize{
Timelike form factor $G$ for the $\Omega^-$.
We present the full result in the left panel
(including $G_{E2}$ and $G_{M3}$).
In the right panel, we drop the higher order multipoles  
($G_{E2}$ and $G_{M3}$).
The model from Ref.~\cite{Omega2}
predicts a large magnitude for $G_{M3}$.
The timelike data support the estimates with a much smaller $G_{M3}$.
Data are from Ref.~\cite{Dobbs17a}. 
See also caption of Fig.~\ref{figLambda}. 
}}
\label{figOmega}
\end{figure*}

CLOE-c provided the first measurements 
of the $\Omega^-$ form factors for nonzero $q^2$~\cite{Dobbs14a,Dobbs17a}.
Our results for the $\Omega^-$ form factors
are very important, because theoretical studies 
of the  $\Omega^-$ are scarce due to its unstable character.
Fortunately, for the $\Omega^-$, lattice QCD 
simulations at the physical point (i.e.~physical strange quark mass) 
exist~\cite{Alexandrou10}.
Since those simulations are at the physical point 
and the meson cloud contamination (kaon cloud) is expected to be small 
due to the large kaon mass,
the lattice QCD data may be considered 
to describe the physical $\Omega^-$.

We consider a model where the $\Omega^-$ 
is described by a dominant $S$ state and 
two different $D$-state components:
one with total quark spin $1/2$, another 
with the total quark spin $3/2$~\cite{NDeltaD,Omega2}.
Each radial wave function is parametrized 
by a unique momentum-range parameter.
The number of free parameters is then five: two  
$D$-state mixture coefficients and three momentum-range parameters.

Our model for the $\Omega^-$~\cite{Omega2} was calibrated by 
the  $\Omega^-$ lattice QCD data from Ref.~\cite{Alexandrou10}.
The free parameters of the radial wave functions and 
$D$-state mixture coefficients
of our model were adjusted by the lattice QCD results for the form factors  
$G_{E0}$, $G_{M1}$, and $G_{E2}$ for $Q^2 < 2$ GeV$^2$.
The model was then used to estimate the functions $G_{E2}$ 
(electric qudrupole form factor) and $G_{M3}(Q^2)$ 
(magnetic octupole form factor).

In the case of the electric quadrupole form factor ($G_{E2}$),
one obtains a consistent description of the lattice QCD data,
which allows the determination of the electric quadrupole moment
from $G_{E2}(0) = 0.680 \pm 0.012$.
For $G_{M3}$, however, the lattice QCD simulations  
are restricted to the result for $Q^2=0.23$ GeV$^2$,
$G_{M3} = 1.25 \pm 7.50$~\cite{Boinepalli09,Omega2}
(which has a significant errorbar).
From the form factors $G_{E0}$, $G_{M1}$, $G_{E2}$ and $G_{M3}$,
we calculate the function $G$ based on the results in Appendix~\ref{appSpin32}.
The results are presented in the left panel of Fig.~\ref{figOmega}.

Our estimates for the 
electromagnetic form factors in the timelike region of $\Omega^-$ 
should be taken with caution, since the model 
used for the radial wave functions was not chosen 
in order to describe the large-$Q^2$ region 
but rather fitted to the $Q^2 < 2 $ GeV$^2$ data.
For that reason, the falloff of $G_{E0}$ and $G_{M1}$ at large $Q^2$
is determined by the $1/Q^6$ behavior, 
and not by the falloff of pQCD ($1/Q^4$).

From  Fig.~\ref{figOmega} (left panel),
we conclude that our results for $G$ overstimate the data.
In order to understand this result, 
we examine the magnitude of the higher multipole form factors 
$G_{E2}$ and $G_{M3}$.
If we drop these contributions we obtain 
the results presented in the right panel of Fig.~\ref{figOmega}.
In this case,  we observe a close agreement with the data.
From this analysis, we can conclude that the deviation from the data 
comes from the form factors $G_{E2}$ and $G_{M3}$.
We have confirmed that it is  the  function  $G_{M3}$ that originates 
a contribution that makes the total results differ from the data.
Our model gives $G_{M3}(0) \simeq 15.5$.
The result presented in the right panel of Fig.~\ref{figOmega}
is more compatible with $G_{M3}(0) \approx 1$.
We then conclude that the timelike data 
are more consistent with a small magnitude for the function $G_{M3}$.

The value of $G_{M3}(0)$ has been estimated 
based on different frameworks.
Light front QCD sum rules predict $G_{M3}(0)=64.3 \pm 16.1$~\cite{Aliev09}.
Estimates based on a non-covariant quark model 
gives $G_{M3}(0)=48.2$ for a symmetric $SU_F(3)$ quark model 
and $G_{M3}(0)=12.1$ when the symmetry is broken~\cite{Buchmann08}. 
Our estimate is then close to the lower estimate for  $G_{M3}(0)$,
and it is more consistent with the estimate that breaks $SU_F(3)$.
The timelike data, however, seem to indicate that $G_{M3}(0)$ may be even smaller.

It is worth noting that the function $G_{M3}$ is, at the moment, 
poorly estimated. On the contrary, 
the functions $G_{E0}$, $G_{M1}$, and $G_{E2}$, 
are well determined by the lattice QCD data.
The present result suggests 
the need for a determination of  $G_{M3}$ 
by a combined study of more accurate lattice QCD data 
with the very recent timelike region data for $G$ 
in the region $q^2 \approx 16$ GeV$^2$~\cite{Dobbs17a}.
In future studies, the expected pQCD falloff 
of the form factors for very large $q^2$ 
should also be taken into account.

\section{Outlook and conclusions}
\label{secConclusions}

A relativistic quark model which was successful in the description 
of the baryon electromagnetic form factors 
in the spacelike region was extended to  the timelike region. 
Our $SU_F(3)$ model
provides a fair description of 
the data both in the spacelike and timelike regions.

The extension of the model from the spacelike 
into the timelike regions uses asymptotic 
reflection symmetry relations connecting the 
electromagnetic elastic form factors in the two different regions.
The theoretical uncertainty in our predictions 
for the timelike region was presented.
An important conclusion is that the measured data 
are consistent with the asymptotic relations of
Eqs.~(\ref{eqGMnew})--(\ref{eqGEnew}), 
originated from general principles as unitarity and analyticity. 
Finite corrections for $q^2$ still have a role in the strength 
of the form factors for $q^2=10$--30 GeV$^2$, 
since within this range the differences between the results 
obtained from $G_l^{\rm SL} (-q^2)$,
$G_l^{\rm SL} (2M_B^2 -q^2)$, and $G_l^{\rm SL} (4 M_B^2 -q^2)$ 
($l=M,E$) show that the strict $q^2 \rightarrow \infty$ limit is not yet 
attained numerically within that region.
On the other hand, the fact that the data 
are within the theoretical uncertainty of our model
seems to indicate that the 
reflection symmetry center point 
is inside the unphysical region $]0,4 M_B^2[$, 
where $M_B$ is the baryon mass.

Our model leads to the correct pQCD asymptotic power 
law behavior of the electromagnetic form factors.
But an important conclusion of this work  is that the pQCD limit onset
$G \propto 1/q^4$ is way above the region where 
the reflection symmetry relations are valid.
We found that only beyond the region of $q^2$: $30-50$ GeV$^2$, 
the pQCD power law was observed.
This was interpreted as an interplay of the two scales entering the model:
the meson mass scales that determine the quark electromagnetic current, 
and the momentum-range scales determined by the extension of the hyperons.

In the present work, our main focus was on the baryon octet 
since the available data are mostly on that family 
of baryons, and therefore, the comparison with the data 
enabled us to better probe our model in the timelike region.
Our framework can also be applied to all baryons of the decuplet, and
as an example, we presented our results 
for the $\Omega^-$ baryon and compared them
with the new data from CLEO.
Under study is the possible extension 
of the present model to charmed baryons.
By this extension, the model can be applied to
the $e^+ e^- \to \Lambda_c^+ \bar \Lambda_c^+ $ process   
to estimate the $\Lambda_c^+$ timelike electromagnetic form factors, 
which were recently measured at BES-III~\cite{Ablikim18b}.

\begin{acknowledgments}
G.R.~was supported by the Funda\c{c}\~ao de Amparo \`a
Pesquisa do Estado de S\~ao Paulo (FAPESP):
Project No.~2017/02684-5, Grant No.~2017/17020-BCO-JP.
M.T.~Pe\~na was supported in part by Funda\c c\~ao para a Ci\^encia 
e a Tecnologia (FCT) Grant No.~CFTP-FCT (UID/FIS/00777/2015).
K.T.~was supported by Conselho Nacional de Desenvolvimento 
Cient\'{i}fico e Tecnol\'ogico (CNPq, Brazil), Processes No.~313063/2018-4 
and No. 426150/2018-0, and FAPESP Process No. 2019/00763-0, 
and his work was also part of the projects, 
Instituto Nacional de Ci\^{e}ncia e 
Tecnologia - Nuclear Physics and Applications 
(INCT-FNA), Brazil, Process No.~464898/2014-5, 
and FAPESP Tem\'{a}tico, Brazil, Process No.~2017/05660-0.
M.~T.~Pe\~na thanks Gernot Eichmann and 
Alfred Stadler for useful discussions.
\end{acknowledgments}

\appendix

\section{$3/2^+$ baryons}
\label{appSpin32}

As discussed in the main text, the relations (\ref{eqSigma1})
and (\ref{eqGeff1}) can be used for $3/2^+$ baryons
if the form factors $G_M$ and $G_E$ 
are expressed as combinations of
electric form factors 
(electric charge $G_{E0}$ and electric quadrupole $G_{E2}$)
and magnetic form factors 
(magnetic dipole $G_{M1}$ and magnetic octupole $G_{M3}$).

According to Ref.~\cite{Korner77}, 
we should use the following replacements:
\ba
& &
|G_E|^2 \to 2 |G_{E0}|^2 + \frac{8}{9} \tau^2 |G_{E2}|^2, \\
& &
|G_M|^2 \to \frac{10}{9}|G_{M1}|^2 + \frac{32}{5} \tau^2 |G_{M3}|^2,
\ea
where $\tau = \frac{q^2}{4 M_B^2}$.

\setcounter{table}{0}
\renewcommand{\thetable}{B\arabic{table}}

\section{Details of the model}
\label{appCSQM}

\begin{table*}[t]
\begin{center}
\begin{tabular}{l c c c}
\hline
\hline
$B$   & $\ket{M_A}$  & &  $\ket{M_S}$  \\ 
\hline
$p$     &    $\sfrac{1}{\sqrt{2}} (ud -du) u$
 & &
 $\sfrac{1}{\sqrt{6}} \left[
        (ud + du) u - 2 uu d \right]$  \\
$n$     &  $\sfrac{1}{\sqrt{2}} (ud -du) d$ & &
 $-\sfrac{1}{\sqrt{6}} \left[
         (ud + du) d - 2 ddu \right]$  \\[.3cm]
$\Lambda^0$ &  $\sfrac{1}{\sqrt{12}}
\left[
s (du-ud) - (dsu-usd) -2(du-du)s
\right]$
& &
$\sfrac{1}{2}
\left[ (dsu-usd) + s (du-ud)
\right]$ \\[.3cm]
$\Sigma^+$  & 
$\sfrac{1}{\sqrt{2}} (us -su) u $ 
& &  
$\sfrac{1}{\sqrt{6}} \left[(us + su) u - 2 uu s \right]$ 
\\
$\Sigma^0$ &
$\sfrac{1}{2}
\left[ (dsu+usd) -s (ud+du)
\right]$ 
& &
$\sfrac{1}{\sqrt{12}} \left[
s (du+ud) +(dsu+usd) -2(ud+du)s
\right]$ \\
$\Sigma^-$ &   $\sfrac{1}{\sqrt{2}} (ds -sd) d$  & &
$\sfrac{1}{\sqrt{6}}\left[ (sd + ds) d - 2 dd s \right]$  \\[.3cm]
$\Xi^0$ &     $\sfrac{1}{\sqrt{2}} (us -su) s$  & &
 $-\sfrac{1}{\sqrt{6}} \left[(ud + du) s - 2 ss u\right]$  \\
$\Xi^-$ &  
$\sfrac{1}{\sqrt{2}} (ds -sd) s$  & &
$-\sfrac{1}{\sqrt{6}} \left[(ds + sd) s - 2 ss d\right]$ \\
\hline
\hline
\end{tabular}
\end{center}
\caption{Flavor wave functions of the octet baryons.}
\label{tablePHI}
\end{table*}

Below we describe some  details of the model, 
including the spin and flavor wave 
functions, the explicit form of the quark form factors,
the parameters of the model, and the values of the normalization factors
due to the inclusion of pion cloud contributions.

\subsection{Baryon wave functions}

In the covariant spectator quark model 
the spin states associated with Eq.~(\ref{eqPsiB})
are represented by~\cite{Octet1,Nucleon}
\ba
\phi_S^0 = u_B(P,s),
\hspace{.2cm}
\phi_S^1 = - (\varepsilon_\lambda^\ast)_\alpha (P) U_B^\alpha(P,s),
\label{eqPsiB2}
\ea
where $u_B$ is the Dirac spinor of the baryon,
$s$ is the baryon spin projection,
$\lambda$ represents the polarization of the diquark, and
\ba
U_B^\alpha (P,s) = 
\frac{1}{\sqrt{3}} \gamma_5 \left(
\gamma^\mu - \frac{P^\alpha}{M_B}  \right) u_B(P,s).
\ea

The wave function described by Eqs.~(\ref{eqPsiB}) and ~(\ref{eqPsiB2})
generalize the nonrelativistic wave function 
in a covariant form~\cite{Nucleon}.
The flavor states $\left| M_S\right>$ 
(mixed symmetric) and  $\left| M_A\right>$ 
(mixed antisymmetric)  for all octet baryons 
are presented in Table~\ref{tablePHI}.
Although the results for the nucleon are not 
discussed in the present work, we include  
the proton and neutron states for completeness.


\subsection{Quark form factors}
\label{secQFF}

To parametrize the quark current (\ref{eqJq}),
we adopt the structure inspired by the VMD mechanism as in Refs.~\cite{Nucleon,Omega}:
\ba
& &
\hspace{-1cm}
f_{1 \pm} = \lambda_q
+ (1-\lambda_q)
\frac{m_v^2}{m_v^2+Q^2} + c_\pm \frac{M_h^2 Q^2}{(M_h^2+Q^2)^2},
\nonumber \\
& &
\hspace{-1cm}
f_{1 0} = \lambda_q
+ (1-\lambda_q)
\frac{m_\phi^2}{m_\phi^2+Q^2} + c_0 \frac{M_h^2 Q^2}{(M_h^2+Q^2)^2},
\nonumber \\
& &
\hspace{-1cm}
f_{2 \pm} = \kappa_\pm
\left\{
d_\pm  \frac{m_v^2}{m_v^2+Q^2} + (1-d_\pm)
\frac{M_h^2 }{M_h^2+Q^2} \right\}, \nonumber \\
& &
\hspace{-1cm}
f_{2 0} = \kappa_s
\left\{
d_0  \frac{m_\phi^2}{m_\phi^2+Q^2} + (1-d_0)
\frac{M_h^2}{M_h^2+Q^2}  \right\},
\label{eqQff}
\ea
where $m_v,m_\phi$ and $M_h$ are the masses, respectively,
corresponding to the light vector meson $m_v \simeq m_\rho \simeq m_\omega$,
the $\phi$ meson (associated with an $s \bar s$ state),
and an effective heavy meson
with mass $M_h= 2 M_N$ to represent
the short-range phenomenology.
The parameter $\lambda_q$ is determined 
by the study of deep inelastic scattering~\cite{Nucleon}.
The relation  between the quark anomalous magnetic moments 
$\kappa_u$ and $\kappa_d$ is 
$\kappa_+ = 2 \kappa_u - \kappa_d$ and 
$\kappa_- = \frac{1}{3}(2 \kappa_u + \kappa_d)$.

We consider the parametrization from Refs.~\cite{Nucleon,Octet1}
in the study of the nucleon and decuplet systems, 
except for the quark anomalous magnetic moments $\kappa_+$ and $\kappa_-$.
Those coefficients are re-adjusted in our study 
of the octet baryon electromagnetic form factors 
in order to take into account the pion cloud effects.
The value of $\kappa_s$ is fixed by the magnetic 
moment of the $\Omega^-$.

The parametrization from Eq.~(\ref{eqQff}) 
for the three sectors 
includes the quark anomalous magnetic moments  (three parameters) 
and six extra parameters.
Since the quark anomalous magnetic moments 
can be fixed independently by the proton, the neutron and the $\Omega^-$ magnetic moments,
we have six parameters to adjust.
We reduce the number to five using the 
isospin symmetry for the Pauli form factor ($d_+=d_-$).
All parameters of the quark current are included in Table~\ref{tabCurrent}.
The results in boldface indicate the values fixed 
by the study of the octet baryons with pion cloud dressing.

\begin{table}[tb]
\begin{center}
\begin{tabular}{c c  c c}
\hline
\hline
$a$ & $\kappa_a$ & $c_a$  & $d_a$  \\
\hline
$+$ &  {\bf 1.462}   &   4.160  & $-0.686$ \\   
$-$ &  {\bf 1.756}  &   1.160 &  $-0.686$ \\  
$0$ &       1.462  &   4.427 & $-1.860$ \\   
\hline
\hline
\end{tabular}
\end{center}
\caption{\footnotesize  
Parameters associated with the quark current. 
In this notation $\kappa_s  = \kappa_0$.
For $\lambda_q$, we use $\lambda_q =1.22$~\cite{Nucleon}.}
\label{tabCurrent}
\end{table}

\subsection{Octet baryon form factors}
\label{appOctet}

The valence quark contributions to the octet baryon 
electromagnetic form factors are then determined by the combination of the 
quark form factors and the parametrization of the radial wave functions,
according to Eqs.~(\ref{eqF1}) and  (\ref{eqF2}).

The parameters associated with the radial wave functions 
(\ref{eqChiB}) are presented in Table~\ref{tableBeta}.
The coefficients $j_i^{A,S}$ ($i=1,2$) 
depend on the quark form factors according to 
the results from Table~\ref{tablePhiB}. 
Again, the expressions for the nucleon 
are included for completeness.

\begin{table}[t] 
\begin{center}
\begin{tabular}{c c c c}
\hline
\hline
$\beta_1$  & $\beta_2$ & $\beta_3$ & $\beta_4$ \\
\hline
0.0440\spQ & 0.9077\spQ &  0.7634\spQ  & 0.4993\spQ  \\
\hline
\hline
\end{tabular}
\end{center}
\caption{\footnotesize 
Parameters of the radial wave functions (\ref{eqChiB}).
Recall that $\beta_1$ is the global long-range parameter.
The short-range parameters are $\beta_2$ (nucleon); 
$\beta_3$ ($\Lambda$ and  $\Sigma$), 
and $\beta_4$ ($\Xi$).}
\label{tableBeta}
\end{table}

\begin{table}[t] 
\begin{center}
\begin{tabular}{l c c}
\hline
\hline
$B$   & $j_i^S$  &   $j_i^A$  \\
\hline
$p$     & $\sfrac{1}{6} (f_{i+}-f_{i-}) $ &
        $\sfrac{1}{6} (f_{i+}+3 f_{i-})  $ \\
$n$     & $\sfrac{1}{6} (f_{i+}+  f_{i-}) $ &
        $\sfrac{1}{6} (f_{i+}-  3 f_{i-})$ \\[.3cm]
$\Lambda^0$ & $\sfrac{1}{6}f_{i+}$ &
 $\sfrac{1}{18} (f_{i+}-  4 f_{i0})$ \\[.3cm]
$\Sigma^+$  & $\sfrac{1}{18}
(f_{i+} + 3 f_{i-} -4 f_{i0}) $ &
 $\sfrac{1}{6} (f_{i+}+3 f_{i-})  $ \\
$\Sigma^0$ &  $\sfrac{1}{36} (2 f_{i+}-  8 f_{i0})$
& $\sfrac{1}{6}f_{i+}$ \\
$\Sigma^-$ & $\sfrac{1}{18}
(f_{i+} - 3 f_{i-} -4 f_{i0}) $ &
        $\sfrac{1}{6} (f_{i+}-  3 f_{i-})           $ \\[.3cm]
$\Xi^0$ & $\sfrac{1}{18} (2 f_{i+} + 6  f_{i-} -2 f_{i0}) $ &
$-\sfrac{1}{3}f_{i0}$ \\
$\Xi^-$ & $\sfrac{1}{18} (2 f_{i+} -6  f_{i-} -2 f_{i0}) $ &
$-\sfrac{1}{3}f_{i0}$ \\
\hline
\hline
\end{tabular}
\end{center}
\caption{\footnotesize 
Mixed symmetric and antisymmetric coefficients for the octet baryons.}
\label{tablePhiB}
\end{table}

\subsection{Pion cloud dressing}
\label{appPionCloud}

In the low-$Q^2$  region, it is necessary to include the 
effects of the pion cloud dressing of the baryons.
In the study of the electromagnetic structure 
of the octet baryons those effects are taken 
into account in an effective way.
There are two main  contributions to take into account:
the contributions associated with the 
photon coupling with the pion, and 
the contributions associated with the photon coupling 
with intermediate octet baryon states.
All these processes can be parametrized based 
on an $SU(3)$ model for the pion-baryon interaction
using five independent couplings and two cutoffs 
(regulate falloff of pion cloud contributions)~\cite{Octet1}.

The main consequence of the inclusion of 
the pion cloud contributions is that the 
estimates of $G_{EB}^B$ and $G_{MB}^B$ from the valence 
quark contributions to the octet baryon form 
factors are modified by the normalization 
of the wave function which combine valence and 
pion cloud contributions ($\delta G_{EB}$ and $\delta G_{MB}$):
\ba
& & G_{EB} = Z_B [ G_{EB}^B + \delta G_{EB}], 
\label{eqGEnorm}\\
& & G_{MB} = Z_B [ G_{MB}^B + \delta G_{MB}]. 
\label{eqGMnorm}
\ea
The explicit expressions for $\delta G_{EB}$ and 
$\delta G_{MB}$ can be found in Refs.~\cite{Octet1,Octet2}.
When we increase $Q^2$ the pion cloud contributions are 
strongly suppressed since they are regulated by 
higher order multipoles with square cutoffs of the order 0.8  
and 1.2 GeV$^2$~\cite{Octet1}.

The parameters associated with the valence quark 
contributions are determined by fits to the 
lattice QCD results for the octet baryon electromagnetic form factors.
The parameters associated with the pion cloud contribution
are fixed by the physical data 
(nucleon electromagnetic form factors and octet baryon magnetic moments).

\begin{table}[b] 
\begin{center}
\begin{tabular}{c |  c c}
\hline
\hline
   $B$  & $a_B$ &  $Z_B$ \\
\hline
\hline
 $N$         &  1   & 0.885 \\ 
 $\Lambda$   &  $\frac{4}{3}\alpha^2$  & 0.941\\
 $\Sigma$  &  $\frac{1}{3}(8 -16 \alpha + 10 \alpha^2)$ & 0.929 \\
 $\Xi$     &  $(1 - 2 \alpha)^2$  & 0.995\\
\hline
\hline
\end{tabular}
\end{center}
\caption{\footnotesize
Coefficients $a_B$ associated with the normalization 
of the octet baryon wave functions,
and numerical result for the normalization constant $Z_B$.
The $SU(6)$ value is $\alpha=0.6$.}
\label{tabZB}
\end{table}

In Eqs.~(\ref{eqGEnorm})--(\ref{eqGMnorm})
the factor $Z_B$ can be written as~\cite{Octet1,Octet2,Octet4}
\ba
Z_B = \frac{1}{1 + 3 a_B B_1},
\ea
where $a_B$ is a coefficient 
determined by the $SU(3)$ symmetry, and $B_1$ is a parameter 
which determines the nucleon normalization ($Z_N$),
based on the normalization  $a_N=1$.

The normalization constant for the nucleon
$Z_N= 1/(1 + 3B_1)$, meaning that the contribution 
from the valence quarks for the proton charge is 
$Z_N$ and the contribution from the pion cloud $3B_1 Z_N$.
One concludes then that the relative pion cloud 
contribution to the proton electric form factor is $3B_1$,
which implies that the normalization $Z_N$ can be determined 
by the estimate of the pion cloud contribution 
based on the comparison between 
the valence quark contributions and the data, 
and vice versa $B_1 = \frac{1}{3}\frac{1- Z_N}{Z_N}$.
The values of $Z_B$ based on the value for $B_1$
and $a_B$ are presented in Table~\ref{tabZB}.




\end{document}